\documentclass[11pt,fleqn]{article}
\usepackage{newcent}

\usepackage[T1]{fontenc}
\usepackage[latin9]{inputenc}
\usepackage[a4paper]{geometry}
\usepackage{color}
\usepackage{textcomp}
\usepackage{amsmath}
\usepackage{amssymb}
\usepackage{setspace}
\usepackage{esint}
\usepackage[unicode=true,pdfusetitle,
 bookmarks=true,bookmarksnumbered=true,bookmarksopen=true,bookmarksopenlevel=1,
 breaklinks=false,pdfborder={0 0 1},backref=false,colorlinks=true,pdfpagemode=FullScreen]
 {hyperref}
\usepackage{breakurl}

\makeatletter
\@ifundefined{date}{}{\date{}}

\makeatother

\begin{document}

\title{Hamilton-Jacobi formalism for Podolsky's electromagnetic theory on
the null-plane}

\maketitle
\begin{center}
M. C. Bertin\footnote{mbertin@ufba.br}, B. M. Pimentel\footnote{pimentel@ift.unesp.br},
C. E. Valcárcel\footnote{carlos.valcarcel@ufabc.edu.br}, G. E. R.
Zambrano\footnote{gramos@udenar.edu.co}
\par\end{center}

\begin{center}
\emph{$^{1}$Instituto de Física, Universidade Federal da Bahia,}\\
\emph{Câmpus Universitário de Ondina, 40210-340, Salvador - BA, Brazil.}
\par\end{center}

\begin{center}
\emph{$^{2}$Instituto de Física Teórica, UNESP - São Paulo State
University,}\\
\emph{Caixa Postal 70532-2, 01156-970, São Paulo, SP, Brazil.}
\par\end{center}

\begin{center}
\emph{$^{3}$Centro de Matemática, Computação e Cognição,}\\
\emph{Universidade Federal do ABC, 09210-170, Santo André, SP, Brazil}
\par\end{center}

\begin{center}
\emph{$^{4}$Departamento de Física, Universidad de Nariño,}\\
\emph{Calle 18 Carrera 50, San Juan de Pasto, Nariño, Colombia.\thispagestyle{empty}}
\par\end{center}
\begin{abstract}
\begin{singlespace}
We develop the Hamilton-Jacobi formalism for Podolsky's electromagnetic
theory on the null-plane. The main goal is to build the complete set
of Hamiltonian generators of the system, as well as to study the canonical
and gauge transformations of the theory.\end{singlespace}

\end{abstract}
\tableofcontents{}

\section{Introduction\label{sec:Introduction}}

In this paper, we analyse the null-plane canonical structure of the
generalised Podolsky's electrodynamics \emph{via} the Hamilton-Jacobi
(HJ) formalism for singular systems. Since we are dealing with three
separate subjects, each one deserving proper treatment by itself,
we separate the work in three main parts.

In the first part, section \ref{sec:The-HJ-formalism}, it is presented
the general Hamilton-Jacobi theory for singular and higher-order lagrangian
systems, as it is the case of Podolsky's theory. The HJ formalism
is well known from classical mechanics to be a road for the study
of integrability of classical systems, but its fundamental role in
dynamical systems was discovered only after Carathéodory's work \cite{Caratheodory}
on variational principles and first-order partial differential equations
(PDEs). Carathéodory built the HJ theory directly from Hamilton's
principle, showing that it is actually the theory that relates first-order
PDEs, first-order ordinary differential equations (ODEs), and lagrangian
variational problems. This unifying point of view is called the \emph{complete
figure of the variational calculus}, and congregates all analytic,
algebraic, and geometric pictures of mechanics \cite{Hund}.

Considering singular systems, which are theories whose Lagrange function
has singular Hessian matrices, the HJ theory does not exhibit the
same fundamental problems seen when the purely hamiltonian picture
is considered \cite{Dirac,Bergmann}. Carathéodory's complete figure
is still valid, and a general treatment of constraints is naturally
emergent \cite{Guler-HJ}. In the HJ theory, canonical constraints
are seen as a set of Hamilton-Jacobi partial differential equations
(HJPDE), and the integrability of this set is achieved by an integrability
theorem, leading to Frobenius' integrability conditions \cite{NICS,ICHJF}.
Advantages of the HJ approach come from the fact that it provides
a full theoretical stage to the canonical formulation of singular
systems, rather than a consistency construction. One of the advantages
is the absence of the so called Dirac's conjecture. Also, gauge fixing
is not required for equivalence between the canonical and lagrangian
descriptions at the classical level. Moreover, second and higher derivative
lagrangian systems were treated \emph{via} the HJ formalism in \cite{general HJ}.
High-order theories were introduced by Ostrogradsky \cite{Ostrogradski},
and have been used since for many purposes. Among them, we may cite
developments in high-order gauge theories \cite{Cuzinatto}, attempts
to solve the problem of renormalisation of the gravitational field
\cite{Stelle - Querella}, and recently quantum gravity massive theories
\cite{NMG}. 

Podolsky's theory has developed by Bopp \cite{Bopp}, and independently
by Podolsky and Schwed \cite{Podolsky} as a second-order gauge theory
for the electromagnetic field, in order to treat the $r^{-1}$ dependency
in the electrostatic potential. Because of this behaviour, the energy
necessary to assemble a system of point charges has infinite contribution
of the charge's self-energy. This problem is related to divergences
in the infrared sector of the quantum electrodynamics (QED), as well
to the problem of infinite vacuum polarisation current. The theory
proposed is described by the lagrangian density
\begin{equation}
L=-\frac{1}{4}F_{\mu\nu}F^{\mu\nu}+\frac{1}{2}a^{2}\partial_{\lambda}F^{\mu\lambda}\partial^{\gamma}F_{\mu\gamma},\label{eq:1.1}
\end{equation}
with $F_{\mu\nu}=\partial_{\mu}A_{\nu}-\partial_{\nu}A_{\mu}$ as
the components of the electromagnetic tensor field, and $a$ being
a parameter with dimension of the inverse of mass. The second-order
derivative term in \eqref{eq:1.1} results in a well defined electrostatic
potential for $r=0$, so the self-energy contribution may be computed,
it is just proportional to $q^{2}/a$ for each charge. The theory
may be interpreted as an effective theory for short distances \cite{Frenkel},
as a way to get rid of the problems related to the $r=0$ singularity
in QED. In this case, the parameter $a$ becomes a cut-off distance
for the theory. When $r\gg a$, the theory would become experimentally
indistinguishable from Maxwell's. In this point of view, $a$ is linked
to an effective radius for the electron.

If interpreted as a fundamental theory, the parameter $a$ is related
to a sector of massive photons, which can be seen by the dispersion
relation 
\begin{equation}
p^{2}\left(p^{2}-m_{\gamma}^{2}\right)=0,\,\,\,\,\,\,\,\,\,\,\,\,p^{2}\equiv p_{\mu}p^{\mu},\,\,\,\,\,\,\,\,\,\,\,\,m_{\gamma}^{2}\equiv1/a^{2},\label{eq:1.2}
\end{equation}
taken from the generalised wave equation, result of the field equations
of the lagrangian \eqref{eq:1.1}. This relation indicates two kinds
of photons, with modes $p^{2}=0$ e $p^{2}=m_{\gamma}^{2}$ respectively.
The first mode corresponds to massless photons, the second mode is
linked to photons with mass parameter $m_{\gamma}$. Massive photons
are not observed in nature, so it is not generally believed that Podolsky's
theory is fundamental. Yet, experimental attempts to provide upper
limits to $m_{\gamma}$ are taken from time to time \cite{Williams}.
The best result, however, is given by Luo \emph{et al} \cite{Luo},
which has $m_{f}<2.1\times10^{-51}g$.

In section \ref{sec:The-null-plane-dynamics} we discuss forms of
relativistic dynamics, specially how to project generators of the
Poincaré group when a given form of dynamics is defined in Minkowski
space-time. The first attempts of quantisation of Podolsky's field
were made in instant-form, where time is defined by the $x^{0}$ axis
as the evolution parameter. There are five different forms of such
hamiltonian dynamics, each one related to different decompositions
of the Poincaré group \cite{rel.dyn.}. The dynamics on the null-plane,
also called the front-form dynamics, is the hamiltonian dynamics of
fields over a null-plane $x^{0}+x^{3}=cte$. The evolution parameter
is chosen to be the coordinate $x^{+}\equiv1/\sqrt{2}\left(x^{0}+x^{3}\right)$,
where the classical (quantum) evolution of the system is given by
the definition of appropriate fundamental ``equal-time'' brackets
(commutators), defined on a null-plane of constant $x^{+}$, plus
a special set of initial-boundary data.

There are some good reasons, both physical and mathematical, to analyse
field theories on the null-plane. One of them is the fact that this
kind of dynamics usually reduces the number of independent degrees
of freedom necessary to describe a field \cite{Huszar,Steinhardt}.
This is closely related to the fact that the stability group of the
Poincaré group in front-form, which is the sub-group of transformations
that relates field configurations in a single surface $x^{+}=cte$,
has seven generators, one more than the six kinematic generators in
instant-form. Besides, the algebra of these kinematic generators takes
its simplest form in front-form dynamics. For some important systems
this feature is responsible for a complete separation of physical
degrees of freedom, resulting in a clean and excitation-free quantum
vacuum.

We let the application of the null-plane HJ formalism for section
\ref{sec:Podolsky's-electromagnetic-theor}. The canonical structure
of Podolsky's theory was already studied in \cite{Galvao}, in instant-form,
and in \cite{Pod1} on the null-plane, both using Dirac's method.
However, these papers had the goal of analysing proper gauge conditions
for the theory, in order to clarify the physical degrees of freedom.
Our main purpose here, using the HJ theory, is to obtain a complete
set of involutive generators, or hamiltonians. We see that the presence
of a non-involutive subset of constraints demands modification of
the dynamics of the system, with the introduction of generalised brackets
(GBs). With these brackets, we are able to find the complete set of
generators, which close a Lie algebra with the GBs, assuring integrability
as stated by Frobenius' theorem. On the other hand, any complete set
of hamiltonians generate canonical transformations, so we also present
a way of defining these transformations and relate them to the gauge
transformations of Podolsky's lagrangian. This is done without any
mention to Dirac's conjecture, or even the necessity of ad-hoc methods,
such as the case of Castellani's procedure \cite{Castellani}. In
section \ref{sec:Final-remarks} we present our final remarks.

\section{The HJ theory for singular systems and higher-order actions\label{sec:The-HJ-formalism}}

\subsection{Higher-order theories\label{subsec:HO theories}}

Let us start with a system described by $n$ generalised coordinates
$q^{i}$. In general, the coordinates should be at least of class
$C^{2k}$ in a time parameter $\tau$, but for now we may define the
variables $q_{I}^{i}$ as the $I$-th derivative of $q^{i}$ with
respect to $\tau$, where $I=\left\{ 0,1,\cdots,k\right\} $ for a
given $k\in\mathbb{N}$. A high-order lagrangian theory, in this case
of order $k$, is a theory described by the action 
\begin{equation}
A\left[\gamma\right]=\int_{\gamma}d\tau L\left[\tau,q_{I}^{i}\left(\tau\right)\right],\label{eq:2.1}
\end{equation}
which is a functional of curve segments $\gamma:q^{i}=q^{i}\left(\tau\right)$
in the space of generalised coordinates.

Supposing $\delta\tau\equiv\bar{\tau}-\tau$ and $\delta q^{i}\equiv\bar{q}^{i}\left(\bar{\tau}\right)-q^{i}\left(\tau\right)$
general first-order variations of the time and coordinates, we have
the first-order variation, 
\begin{equation}
\delta A=\int d\tau\left[\frac{\delta L}{\delta q^{i}}\left(\delta-\delta\tau\frac{d}{d\tau}\right)q^{i}+\frac{d}{d\tau}\left(\frac{\delta L}{\delta q_{\left(I+1\right)}^{i}}\delta q_{I}^{i}-H\delta\tau\right)\right].\label{eq:2.2}
\end{equation}
In \eqref{eq:2.2}, the derivatives 
\begin{equation}
\frac{\delta}{\delta q^{i}}\equiv\sum_{I=0}^{k}\left(-1\right)^{I}\frac{d^{I}}{d\tau^{I}}\frac{\partial}{\partial q_{I}^{i}},\label{eq:2.3}
\end{equation}
are known as the Lagrange derivatives, and 
\begin{equation}
H\equiv\frac{\delta L}{\delta q_{\left(I+1\right)}^{i}}\,q_{\left(I+1\right)}^{i}-L,\label{eq:2.4}
\end{equation}
is defined as the hamiltonian function of the system. Summation is
implicit for all repeated indexes, except in some few expressions,
as \eqref{eq:2.3}. If the action $A\left[\gamma\right]$ has an extreme
configuration $\gamma_{0}$, this configuration is a solution of the
Euler-Lagrange (EL) equations 
\begin{equation}
\frac{\delta L}{\delta q^{i}}=0,\label{eq:2.5}
\end{equation}
which are the equations of motion of the action \eqref{eq:2.1}.

\subsection{Equivalent lagrangians\label{sub:Equivalent-Lagrangians}}

The HJ formalism in Carathéodory's point of view \cite{Caratheodory}
requires a way of defining equivalent actions of the functional \eqref{eq:2.1}.
This is done by supposing a point-transformation $q^{i}\rightarrow q'^{i}$
with generating function $S\left(\tau,q_{I'}^{i}\right)$. With this
transformation, the new lagrangian is related to the old one by $L'=L-dS/d\tau$.
Here we introduce another index $I'=\left\{ 0,1,\cdots,k-1\right\} $,
which will be useful ahead. Adding a total derivative to the lagrangian
does not change the equations of motion, so if $\gamma_{0}$ is an
extreme configuration of the action $\int Ld\tau$, it is also an
extreme configuration of the transformed action $\int L'd\tau$. If
this point-transformation leads to $L'=0$ for $\gamma=\gamma_{0}$,
and $L'>0$ or $L'<0$ for any $\gamma\neq\gamma_{0}$ in a close
neighbourhood of $\gamma_{0}$, $A\left[\gamma\right]$ has, respectively,
a local minimum $\left(L'>0\right)$ or a local maximum $\left(L'<0\right)$
in $\gamma_{0}$. Expanding $L'$ in a Taylor series around $\gamma_{0}$,
we see that the new lagrangian is approximated by a lower-order term
that is quadratic positive (or negative) definite in $\delta q_{I}^{i}$.
The necessary conditions for the extreme configuration become 
\begin{equation}
\left.\frac{\partial L'}{\partial q_{I}^{i}}\right|_{\gamma=\gamma_{0}}=0,\label{eq:2.6}
\end{equation}
which result in 
\begin{equation}
\frac{\partial S}{\partial q_{I'}^{i}}=\frac{\partial L}{\partial q_{\left(I'+1\right)}^{i}}-\frac{d}{d\tau}\left(\frac{\partial S}{\partial q_{\left(I'+1\right)}^{i}}\right),\label{eq:2.7}
\end{equation}
at $\gamma=\gamma_{0}$.

If we take $I'=k-1$, \eqref{eq:2.7} becomes 
\begin{equation}
\frac{\partial S}{\partial q_{\left(k-1\right)}^{i}}=\frac{\partial L}{\partial q_{k}^{i}},\label{eq:2.8}
\end{equation}
For $I'=k-2$ we have 
\[
\frac{\partial S}{\partial q_{\left(k-2\right)}^{i}}=\frac{\partial L}{\partial q_{\left(k-1\right)}^{i}}-\frac{d}{d\tau}\left(\frac{\partial S}{\partial q_{\left(k-1\right)}^{i}}\right).
\]
With \eqref{eq:2.8}, the result is 
\[
\frac{\partial S}{\partial q_{\left(k-2\right)}^{i}}=\frac{\partial L}{\partial q_{\left(k-1\right)}^{i}}-\frac{d}{d\tau}\frac{\partial L}{\partial q_{k}^{i}}=\frac{\delta L}{\delta q_{\left(k-1\right)}^{i}},
\]
where we recognise the Lagrange derivative on the right. Following
this iterative process, the general relations are found to be 
\begin{equation}
\frac{\partial S}{\partial q_{I'}^{i}}=\frac{\delta L}{\delta q_{\left(I'+1\right)}^{i}}.\label{eq:2.9}
\end{equation}
On the other hand, $L'=L-dS/d\tau$ immediately leads to 
\begin{equation}
L-\frac{\partial S}{\partial\tau}-q_{\left(I'+1\right)}^{i}\frac{\partial S}{\partial q_{I'}^{i}}=0.\label{eq:2.10}
\end{equation}
Therefore, the existence of a function $S\left(\tau,q_{I'}^{i}\right)$
that obeys the conditions \eqref{eq:2.9} and \eqref{eq:2.10} is
a necessary condition for the existence of an extreme configuration
of the action \eqref{eq:2.1}.

\subsection{The HJ equation\label{sub:The-HJ-equation}}

The HJ formalism may be reached if we make of eq. \eqref{eq:2.10}
a PDE for the function $S$. In principle, we expect solving this
problem with expressions of the type $q_{\left(I'+1\right)}^{i}=\phi_{\left(I'+1\right)}^{i}\left(\tau,q_{I'}^{i},\partial S/\partial q_{I'}^{i}\right),$
that would be taken from \eqref{eq:2.9}. This, however, is not the
case. Let us take again the expression \eqref{eq:2.9} for $I'=k-1$:
\[
\frac{\partial S}{\partial q_{\left(k-1\right)}^{i}}=\frac{\delta L}{\delta q_{k}^{i}}=\frac{\partial L}{\partial q_{k}^{i}}=\psi_{i}\left(\tau,q_{I'}^{i},q_{k}^{i}\right),
\]
where $\psi$ is function of $\tau$ and the variables $q_{I}^{i}$.
This equation can indeed be inverted to produce the expression 
\begin{equation}
q_{k}^{i}=\phi_{k}^{i}\left(\tau,q_{I'}^{i},\frac{\partial S}{\partial q_{\left(k-1\right)}^{i}}\right),\label{eq:2.11}
\end{equation}
if the Hessian condition 
\begin{equation}
\det\left(\frac{\partial\psi_{i}}{\partial q_{k}^{j}}\right)=\det\left(\frac{\partial^{2}L}{\partial q_{k}^{i}\partial q_{k}^{j}}\right)\neq0\label{eq:2.12}
\end{equation}
is satisfied.

However, eq. \eqref{eq:2.9} for $I'=k-2$ yields 
\begin{equation}
\frac{\partial S}{\partial q_{\left(k-2\right)}^{i}}=\frac{\partial L}{\partial q_{\left(k-1\right)}^{i}}-\frac{\partial^{2}L}{\partial\tau\partial q_{k}^{i}}-q_{\left(I'+1\right)}^{j}\frac{\partial^{2}L}{\partial q_{I'}^{j}\partial q_{k}^{i}}-q_{\left(k+1\right)}^{j}\frac{\partial^{2}L}{\partial q_{k}^{j}\partial q_{k}^{i}},\label{eq:2.13}
\end{equation}
where a derivative superior to $k$ appears in the last term. This
equation provides 
\begin{equation}
q_{\left(k+1\right)}^{j}=\phi_{\left(k+1\right)}^{j}\left(\tau,q_{I'}^{i},\frac{\partial S}{\partial q_{\left(k-1\right)}^{i}},\frac{\partial S}{\partial q_{\left(k-2\right)}^{i}}\right),\label{eq:2.14}
\end{equation}
if, again, \eqref{eq:2.12} is satisfied. All orders have the same
behaviour. Particularly, the lowest order $I'=0$ results in 
\[
q_{\left(2k\right)}^{i}=\phi_{\left(2k\right)}^{i}\left(\tau,q_{I'}^{i},\frac{\partial S}{\partial q_{I'}^{i}}\right).
\]
Therefore, the best we can do is to invert \eqref{eq:2.9} to obtain
expressions for the variables of the theory from order $k$ to $2k$.
The condition for that is the Hessian matrix 
\begin{equation}
W_{ij}\equiv\frac{\partial^{2}L}{\partial q_{k}^{i}\partial q_{k}^{j}}\label{eq:2.15}
\end{equation}
to be non-singular.

This is not a problem if we consider the variables $q_{I'}^{i}$ independent
coordinates of a configuration space $\mathbb{Q}$. This interpretation
is also suggested by the end-point term in \eqref{eq:2.2}, which
defines the structure of the canonically conjugated variables of the
theory: that sum is a linear combination of independent variations
$\delta q_{I'}^{i}$ only up to order $I=k-1$, since $L$ depends
of the coordinates up to order $k$. In this case, \eqref{eq:2.10}
is just written by 
\begin{equation}
\frac{\partial S}{\partial\tau}+\frac{\partial S}{\partial q_{\left(I'-1\right)}^{i}}q_{I'}^{i}+\frac{\partial S}{\partial q_{\left(k-1\right)}^{i}}\phi_{k}^{i}-L\left[\tau,q_{I'}^{i},\phi_{k}^{i}\right]=0,\label{eq:2.16}
\end{equation}
and no other equation is needed.

Now let us observe the hamiltonian function defined in \eqref{eq:2.4},
which can be written using \eqref{eq:2.9} by 
\begin{equation}
H=\frac{\delta L}{\delta q_{I'}^{i}}\,q_{I'}^{i}+\frac{\delta L}{\delta q_{k}^{i}}\,\phi_{k}^{i}-L=\frac{\partial S}{\partial q_{\left(I'-1\right)}^{i}}\,q_{I'}^{i}+\frac{\partial S}{\partial q_{\left(k-1\right)}^{i}}\,\phi_{k}^{i}-L\label{eq:2.17}
\end{equation}
for $I'>0$. Substituting \eqref{eq:2.11}, we may write \eqref{eq:2.16}
as 
\begin{equation}
\frac{\partial S}{\partial\tau}+H\left(\tau,q_{I'}^{i},\frac{\partial S}{\partial q_{I'}^{i}}\right)=0.\label{eq:2.18}
\end{equation}
Eq. \eqref{eq:2.18} is the expected Hamilton-Jacobi equation.

The introduction of canonical variables is straightforward. In the
HJ formalism, the momenta conjugated to the variables $q_{I'}^{i}$
are defined as the components of the gradient of the function $S$,
\begin{equation}
p_{i}^{I'}\equiv\frac{\partial S}{\partial q_{I'}^{i}}=\frac{\delta L}{\delta q_{\left(I'+1\right)}^{i}}.\label{eq:2.19}
\end{equation}
This definition results in the canonical hamiltonian 
\begin{equation}
H_{0}\left(\tau,q_{I'}^{i},p_{i}^{I'}\right)\equiv p_{i}^{\left(I'-1\right)}q_{I'}^{i}+p_{i}^{\left(k-1\right)}\phi_{k}^{i}-L,\label{eq:2.20}
\end{equation}
with all $q_{k}^{i}$ substituted by $\phi_{k}^{i}$ as functions
of the momenta. It also results in the canonical HJ equation 
\begin{equation}
\Phi_{0}\left(\tau,p_{\tau},q_{I'}^{i},p_{i}^{I'}\right)\equiv\pi_{0}+H_{0}\left(\tau,q_{I'}^{i},p_{i}^{I'}\right)=0,\label{eq:2.21}
\end{equation}
where we introduce the notation $\pi_{0}\equiv\partial S/\partial\tau$.

\subsection{The HJ equations for singular systems\label{sub:The-HJ-equations}}

The Hessian condition appears in the HJ formalism as a condition for
the existence of the PDE \eqref{eq:2.18}. But in the case of violation
of the Hessian condition \eqref{eq:2.12}, implying the existence
of constraints, it is still straightforward to define a HJ approach.
Let us suppose the $n\times n$ Hessian matrix $W_{ij}$ to be singular
of rank $m<n$. Then, there is a regular sub-matrix $m\times m$ and
a null-space of dimension $r=n-m$, splitting the configuration space
in two subspaces: the space of the variables $q_{I'}^{a}$, for $a=\left\{ 1,\cdots,m\right\} $,
which will be called $\mathbb{Q}_{m}$, and the space of the variables
$t_{I'}^{z}\equiv q_{I'}^{z}$ with $z=\left\{ 1,\cdots,r\right\} $,
which will be called $\Gamma_{r}$. The former are variables belonging
to the regular space of the Hessian, related to the matrix 
\begin{equation}
W_{ab}\equiv\frac{\partial^{2}L}{\partial q_{k}^{a}\partial q_{k}^{b}},\label{eq:2.22}
\end{equation}
which obeys $\det W_{ab}\neq0$. The later are variables belonging
to the null-space of the same Hessian matrix.

In the singular case, eq. \eqref{eq:2.9} produce the $m\cdot k$
relations 
\begin{equation}
q_{\left(k+I'\right)}^{a}=\phi_{\left(k+I'\right)}^{a}\left(\tau,t_{I'}^{z},q_{I'}^{a},p_{a}^{I'}\right),\,\,\,\,\,\,\,\,p_{a}^{I'}\equiv\frac{\partial S}{\partial q_{I'}^{a}},\label{eq:2.23}
\end{equation}
but also the $r\cdot k$ identities 
\begin{equation}
\frac{\partial S}{\partial t_{I'}^{z}}=\frac{\delta L}{\delta q_{\left(I'+1\right)}^{z}}\equiv-H_{z}^{I'}\left(\tau,t_{I'}^{z},q_{I'}^{a},p_{a}^{I'}\right).\label{eq:2.24}
\end{equation}
In canonical form, \eqref{eq:2.24} become 
\begin{equation}
\Phi_{z}^{I'}\left(\tau,t_{I'}^{y},q_{I'}^{a},\pi_{y}^{I'},p_{a}^{I'}\right)\equiv\pi_{z}^{I'}+H_{z}^{I'}=0,\,\,\,\,\,\,\,\,\pi_{z}^{I'}\equiv\frac{\partial S}{\partial t_{I'}^{z}}.\label{eq:2.25}
\end{equation}
Eqs. \eqref{eq:2.25} form a set of $r\cdot k$ canonical constraints,
and also a set of $r\cdot k$ first-order PDEs.

We notice that the canonical hamiltonian function \eqref{eq:2.17}
does not depend on the variables $t_{k}^{z}$, so the definition of
this function is not dependent of the Hessian condition, and the HJ
equation \eqref{eq:2.21} is still valid. Let us introduce the notation
$t_{I'}^{\alpha}=\left(\tau,t_{I'}^{z}\right)$, then $\alpha=\left\{ 0,1,\cdots,r\right\} $.
Since the momentum conjugated to $\tau$ is named $\pi_{0}$, we may
also include these as new variables in the theory as $\pi_{\alpha}^{I'}=\left(\pi_{0},\pi_{z}^{I'}\right)$.
Then, eqs. \eqref{eq:2.21} and \eqref{eq:2.22} may be written in
a unified way, 
\begin{equation}
\Phi_{\alpha}^{I'}\left(t_{I'}^{\beta},q_{I'}^{a},\pi_{\beta}^{I'},p_{a}^{I'}\right)\equiv\pi_{\alpha}^{I'}+H_{\alpha}^{I'}\left(t_{I'}^{\beta},q_{I'}^{a},p_{a}^{I'}\right)=0,\label{eq:2.26}
\end{equation}
where we also use $H_{\alpha}^{I'}=\left(H_{0},H_{z}^{I'}\right)$.
Eqs. \eqref{eq:2.26} form a set of Hamilton-Jacobi first-order partial
differential equations.

\subsection{Integrability and characteristic equations\label{sub:The-characteristic-equations}}

The HJ equations \eqref{eq:2.26} are necessary, but still not sufficient
for the existence of extreme configurations of the action $A\left[\gamma\right]$.
It is still necessary that \eqref{eq:2.26} provides at least one
complete solution for the $S$ function. These equations are generally
a set of $\left(r+1\right)\cdot k$ non-linear coupled PDEs of the
first-order, so we expect that a complete solution contains $\left(r+1\right)\cdot k$
constants of integration related to $\left(r+1\right)\cdot k$ linearly
independent parameters. In other words, we expect that a complete
solution has the form $S=S\left[t_{I'}^{\alpha},q_{I'}^{a}\left(t_{I'}^{\alpha}\right)\right]$
for a set of $\left(r+1\right)\cdot k$ parameters $t_{I'}^{\alpha}$
and a set of $m\cdot k$ variables $q_{I'}^{a}\left(t_{I'}^{\alpha}\right)$.
Of course, this is possible only if $\Phi_{\alpha}^{I'}$ form a set
of $\left(r+1\right)\cdot k$ linearly independent equations. For
the following, there is no need to carry the $I'$ index. We just
write the generators $\Phi_{\alpha}\equiv\Phi_{\alpha}^{I'}$, in
which now $\alpha=0,\cdots,\left(r+1\right)\cdot k$, and use the
compact notation $q^{i}\equiv\left(t_{I'}^{\alpha},q_{I'}^{a}\right)=\left(t^{\alpha},q^{a}\right)$
and $p_{i}\equiv\left(\pi_{\alpha}^{I'},p_{a}^{I'}\right)=\left(\pi_{\alpha},p_{a}\right)$.

The necessary and sufficient conditions for complete integrability
are the Frobenius' integrability conditions \cite{ICHJF} 
\begin{equation}
\left\{ \Phi_{\alpha},\Phi_{\beta}\right\} =C_{\alpha\beta}^{\,\,\,\,\gamma}\Phi_{\gamma},\label{eq:2.27}
\end{equation}
in which $C_{\alpha\beta}^{\,\,\,\,\gamma}$ is a set of structure
coefficients. The brackets are the complete Poisson brackets (PB)
\begin{equation}
\left\{ A,B\right\} \equiv\frac{\partial A}{\partial q^{i}}\frac{\partial B}{\partial p_{i}}-\frac{\partial B}{\partial q^{i}}\frac{\partial A}{\partial p_{i}}.\label{eq:2.28}
\end{equation}
If \eqref{eq:2.27} holds, $\Phi_{\alpha}$ form a complete set of
constraints in involution with PB operation. We name these set involutive
constraints. Therefore, Frobenius' conditions are resumed in the fact
that $\Phi_{\alpha}$ are generators of a Lie algebra with the PB.

If $F\left(q^{i},p_{i}\right)$ is an observable of the complete phase-space,
its dynamics is given by the fundamental differential 
\begin{equation}
dF=\left\{ F,\Phi_{\alpha}\right\} dt^{\alpha},\label{eq:2.29}
\end{equation}
so the dynamics takes place in the complete phase-space of the variables
$\xi^{i}\equiv\left(t_{I'}^{\alpha},q_{I'}^{a},\pi_{\alpha}^{I'},p_{a}^{I'}\right)$,
where $\Phi_{\alpha}^{I'}\left(\xi\right)$ are the generators, and
$t_{I'}^{\alpha}$ are the evolution parameters. This phase-space
is actually degenerate. Let us calculate (now with explicit indexes
$I'$) 
\begin{equation}
dq_{I'}^{a}=\left\{ q_{I'}^{a},\Phi_{\alpha}^{J'}\right\} dt_{J'}^{\alpha}=\left\{ q_{I'}^{a},p_{b}^{K'}\right\} \frac{\partial\Phi_{\alpha}^{J'}}{\partial p_{b}^{K'}}dt_{J'}^{\alpha}.\label{eq:2.30}
\end{equation}
Using the definition \eqref{eq:2.28}, we see that $\left\{ q_{I'}^{a},p_{b}^{K'}\right\} =\delta_{b}^{a}\delta_{I'}^{K'}$,
so 
\begin{equation}
dq_{I'}^{a}=\frac{\partial\Phi_{\alpha}^{J'}}{\partial p_{a}^{I'}}dt_{J'}^{\alpha}.\label{eq:2.31}
\end{equation}
On the other hand, 
\begin{equation}
dp_{a}^{I'}=\left\{ p_{a}^{I'},\Phi_{\alpha}^{J'}\right\} dt_{J'}^{\alpha}=\left\{ p_{a}^{I'},q_{K'}^{b}\right\} \frac{\partial\Phi_{\alpha}^{J'}}{\partial q_{K'}^{b}}dt_{J'}^{\alpha}=-\frac{\partial\Phi_{\alpha}^{J'}}{\partial q_{I'}^{a}}dt_{J'}^{\alpha}.\label{eq:2.32}
\end{equation}
Eqs. \eqref{eq:2.31} and \eqref{eq:2.32} are the generalisation
of Hamilton's equations for singular systems. They are generally called
the characteristic equations of the system \eqref{eq:2.26}. A third
fundamental characteristic equation is given by 
\begin{equation}
dS=\left\{ S,\Phi_{\alpha}^{J'}\right\} dt_{J'}^{\alpha}=p_{a}^{I'}dq_{I'}^{a}+\pi_{\alpha}^{I'}dt_{I'}^{\alpha}-\Phi_{\alpha}^{I'}dt_{I'}^{\alpha},\label{eq:2.33}
\end{equation}
and results in an integral equations for $S$ if \eqref{eq:2.31}
and \eqref{eq:2.32} provide solutions for $p_{a}^{I'}\left(t_{I'}^{\alpha}\right)$
and $q_{I'}^{a}\left(t_{I'}^{\alpha}\right)$. The CE for $t_{I'}^{\alpha}$
and $\pi_{\alpha}^{I'}$ are just identities, which reveals the degenerate
character of the complete phase space.

There is, however, a reduced phase-space which is not degenerate.
It is the space of the variables $\xi^{A}\equiv\left(q_{I'}^{a},p_{a}^{I'}\right).$
In fact, it is shown in \cite{ICHJF} that the sector of the variables
$\left(t_{I'}^{\alpha},\pi_{\alpha}^{I'}\right)$ in the complete
phase space has zero volume element. The reason for that is the fact
that $t_{I'}^{\alpha}$ form a parameter space which is isomorphic
to a complete affine vector space. In other words, the solutions of
the characteristic equations are trajectories in the reduced phase-space,
of the type $\xi^{A}=\xi^{A}\left(t_{I'}^{\alpha}\right)$, parametrised
by the variables $t_{I'}^{\alpha}$. For this reason, we call $\xi^{A}$
the dependent variables, and $t_{I'}^{\alpha}$ the independent variables
of the theory. Usually, the dependent variables are related to the
``true'' degrees of freedom of a physical system.

\subsection{The generalised brackets\label{sub:GB}}

The Frobenius' theorem implies the HJ equations $\Phi_{\alpha}=0$
must be complete and linearly independent. However, in physical examples
this condition is not usually satisfied. If the set $\Phi=\left\{ \Phi_{\alpha}\right\} $
is not integrable, two possible reasons are: (1) there may be other
HJ equations not included in this set, case in which the equations
are not complete, or (2) a subset of $\Phi$ is not linearly independent.
In this case, since a physical system must be integrable, the aim
is to discover a complete set of involutive constraints. For that,
we use the procedure outlined in \cite{NICS}.

If the HJ equations $\Phi_{\alpha}=0$ are valid, Frobenius' condition
$\left\{ \Phi,\Phi\right\} \subset\Phi$ implies the generators are
dynamical invariants, in other words, 
\begin{equation}
d\Phi_{\alpha}=\left\{ \Phi_{\alpha},\Phi_{\beta}\right\} dt^{\beta}=0.\label{eq:2.34}
\end{equation}
Eq. \eqref{eq:2.34} is a set of linear equations for the differential
of the independent variables, 
\begin{equation}
M_{\alpha\beta}dt^{\beta}=0,\label{eq:2.35}
\end{equation}
where we introduced the matrix $M_{\alpha\beta}\equiv\left\{ \Phi_{\alpha},\Phi_{\beta}\right\} $.
Remember that, in our case, this matrix has dimension $\left(r+1\right)\cdot k$.

We suppose the general case in which $M_{\alpha\beta}$ has rank $p\leq\left(r+1\right)\cdot k$.
Then, there is an invertible sub-matrix $M_{xy}$ whose entries are
the PB between a subset $\left\{ \Phi_{x}\right\} \subset\Phi$, with
$x=1,\cdots,p$. In this case, for $\alpha=x$ \eqref{eq:2.35} becomes
\begin{equation}
M_{x\beta}dt^{\beta}=0\,\,\,\implies\,\,\,M_{xy}dt^{y}=-M_{x\alpha'}dt^{\alpha'},\label{eq:2.36}
\end{equation}
in which $\alpha'=1,\cdots,q$ is the index of the null space of $M_{\alpha\beta}$.
Then, \eqref{eq:2.35} can be written as 
\begin{equation}
dt^{x}=-\left(M^{-1}\right)^{xy}\left\{ \Phi_{y},\Phi_{\alpha'}\right\} dt^{\alpha'},\label{eq:2.37}
\end{equation}
where $\left(M^{-1}\right)^{xy}$ is the inverse matrix related to
$M_{xy}$. Therefore, if the set $\Phi$ is not integrable, the former
independent variables $t^{\alpha}\equiv\left(t^{x},t^{\alpha'}\right)$
are not mutually independent, due to the fact that the matrix $M_{\alpha\beta}$
has non-zero rank. In fact, \eqref{eq:2.37} can be written for any
set of non involutive constraints, not only in the case of maximal
rank.

Let us write \eqref{eq:2.35} for $\alpha=\alpha'$, 
\begin{equation}
M_{\alpha'\beta}dt^{\beta}=0\,\,\,\implies\,\,\,M_{\alpha'x}dt^{x}+M_{\alpha'\beta'}dt^{\beta'}=0.\label{eq:2.38}
\end{equation}
Using \eqref{eq:2.37}, we have 
\begin{equation}
\left[\left\{ \Phi_{\alpha'},\Phi_{\beta'}\right\} -\left\{ \Phi_{\alpha'},\Phi_{x}\right\} \left(M^{-1}\right)^{xy}\left\{ \Phi_{y},\Phi_{\beta'}\right\} \right]dt^{\beta'}=0.\label{eq:2.39}
\end{equation}
Now we introduce the generalised brackets (GB) between two observables
$A$ and $B$: 
\begin{equation}
\left\{ A,B\right\} ^{*}\equiv\left\{ A,B\right\} -\left\{ A,\Phi_{x}\right\} \left(M^{-1}\right)^{xy}\left\{ \Phi_{y},B\right\} ,\label{eq:2.40}
\end{equation}
then we may write \eqref{eq:2.38} as 
\begin{equation}
\left\{ \Phi_{\alpha'},\Phi_{\beta'}\right\} ^{*}dt^{\beta'}=0.\label{eq:2.41}
\end{equation}
It is straightforward to show that the GB are good brackets. They
are antisymmetric bilinear differential operators that obey the Jacobi
identity.

Suppose we have chosen a subset $\left\{ \Phi_{x}\right\} \subset\Phi$
that is not of maximal rank. In this case, \eqref{eq:2.41} yields
again two sets of equations, one of them will give rise to another
generalised brackets, since not all linear combinations of the independent
variables was discovered. The other set is just \eqref{eq:2.41} again.
When all linear combinations of the parameters are eliminated by a
final GB, the differentials $dt^{\alpha'}$ in \eqref{eq:2.41} are
truly linearly independent. Then, the integrability conditions are
just given by 
\begin{equation}
\left\{ \Phi_{\alpha'},\Phi_{\beta'}\right\} ^{*}=0.\label{eq:2.42}
\end{equation}

Eqs. \eqref{eq:2.42} may give two results. It may be identically
satisfied for some generators, but it may result in relations between
the phase-space variables, which should be taken as new constraints.
If found, these new HJ equations should be added to the set $\Phi'\equiv\left\{ \Phi_{\alpha'}\right\} $.
Eqs. \eqref{eq:2.41} should be used again, since it is possible that
the new set $\Phi'$ is non-involutive, and new GB may be defined.
This process must be repeated until \eqref{eq:2.42} gives no more
constraints, completing the system of generators, which will happen
if the system is really integrable. In this case, the remaining set
of constraints obeys the conditions 
\begin{equation}
\left\{ \Phi_{\alpha'},\Phi_{\beta'}\right\} ^{*}=C_{\alpha'\beta'}^{\,\,\,\,\,\thinspace\thinspace\thinspace\,\gamma'}\Phi_{\gamma'},\label{eq:2.43}
\end{equation}
this time with the GB. Therefore, if the system is integrable we must
be able to reduce it to a set of complete involutive constraints with
a GB.

With a complete involutive set, the fundamental differential of an
observable $F$ is given by 
\begin{equation}
dF=\left\{ F,\Phi_{\alpha}\right\} dt^{\alpha}=\left\{ F,\Phi_{\alpha'}\right\} ^{*}dt^{\alpha'},\label{eq:2.44}
\end{equation}
with use of \eqref{eq:2.37}. Note that the generators belonging to
the invertible part of the matrix $M_{\alpha\beta}$, the ones that
are not involutive, are now completely eliminated since the GB of
a non involutive constraint with any observable of the phase-space
is identically zero. If a system has non involutive constraints, the
characteristic equations must be calculated from \eqref{eq:2.44}.
Therefore, a subset of non involutive constraints is responsible to
modify the dynamics of the system.

\subsection{Canonical and lagrangian symmetries\label{sub:Canonical-and-Lagrangian}}

If $\Phi$ is a complete set of involutive constraints, it is shown
in \cite{ICHJF} that each member of the set is a generator of an
active canonical transformation on the reduced phase-space, called
a characteristic flow (CF). This is better seen in a geometrical framework:
the symplectic structure of the complete phase-space is a degenerate
2-form $\omega=-d\theta$, where $\theta=dS$ is the fundamental lepagian
1-form defined by the CE \eqref{eq:2.33}. The symplectic form is
degenerate since it can be written as the sum of a regular 2-form
$\omega_{P}=dq^{a}\wedge dp_{a}$, in canonical coordinates, and a
singular 2-form $a=dt^{\alpha}\wedge d\pi_{\alpha}$ when the HJ equations
$\Phi_{\alpha}=0$ are satisfied. In fact, the 2-form $a$ is identically
a zero form when the Frobenius' conditions \eqref{eq:2.43} are satisfied.
The 2-form $\omega_{P}$ defines the symplectic structure of the reduced
phase-space.

A canonical transformation is any change in the phase-space that preserves
its symplectic structure. To relate each $\Phi_{\alpha}$ to a canonical
transformation, we define a vector field $X_{\alpha}\equiv\left\{ \bullet,\Phi_{\alpha}\right\} ^{*}$
to each $\Phi_{\alpha}$ (the $\bullet$ symbol takes the place of
a phase-space observable). If the set $\Phi$ is completely integrable,
the Lie algebra \eqref{eq:2.43} of the generators implies a Lie algebra
of the vector fields $X_{\alpha}$, 
\begin{equation}
\left[X_{\alpha},X_{\beta}\right]=X_{\alpha}X_{\beta}-X_{\beta}X_{\alpha}=f_{\alpha\beta}^{\,\,\,\,\,\,\gamma}X_{\gamma},\label{eq:2.45}
\end{equation}
if the structure coefficients $f_{\alpha\beta}^{\,\,\,\,\,\,\gamma}=-C_{\alpha\beta}^{\,\,\,\,\,\,\gamma}$
are independent of the complete phase-space variables. Supposing \eqref{eq:2.45}
holds, any operator with the form $T\equiv\exp\left(\Delta t^{\alpha}X_{\alpha}\right)$
is a member of the group of canonical transformations if, and only
if, $\left[\omega,T\right]=0$, which actually is the case. Even if
\eqref{eq:2.45} is not obeyed, it is shown that $\left[X_{\alpha},\omega\right]=0$,
which implies the symplectic structure is preserved by the infinitesimal
first-order transformation 
\begin{equation}
\delta\xi=\delta t^{\alpha}X_{\alpha}\xi.\label{eq:2.46}
\end{equation}
We call \eqref{eq:2.46} a characteristic flow generated by the vector
fields $X_{\alpha}$. Note that the CE \eqref{eq:2.31}, \eqref{eq:2.32}
themselves have the form of characteristic flows $d\xi=dt^{\alpha}X_{\alpha}\xi$.
Therefore, the Frobenius' theorem implies each $\Phi_{\alpha}$ generates
its own CF, related to an independent variable $t^{\alpha}$, and
each of these flows is independent of the others. Moreover, a general
CF is a linear combination of the flows generated by each $X_{\alpha}$,
with the form \eqref{eq:2.46}.

A special CF is defined as follows. Suppose $\delta t^{0}=\delta\tau=0$
and $\delta t^{z}=\epsilon^{z}$, where $\epsilon^{z}$ form a set
of LI constant parameters. In this case, 
\begin{equation}
\delta\xi=\epsilon^{z}X_{z}\xi=\left\{ \xi,\Phi_{z}\right\} ^{*}\epsilon^{z}.\label{eq:2.47}
\end{equation}
The transformation \eqref{eq:2.47} is a canonical flow if the generators
$\Phi_{z}$ close themselves a Lie algebra $\left\{ \Phi_{x},\Phi_{y}\right\} ^{*}=C_{xy}^{\,\,\,\,\,\,z}\Phi_{z}$.
On the other hand, the only way to assure this algebra is if it is
restricted to the reduced phase-space, where $\Phi_{\alpha}=0$ holds.
In this case, all algebrae become abelian. The transformation \eqref{eq:2.47},
restricted to the reduced phase-space where $\left\{ \Phi_{x},\Phi_{y}\right\} ^{*}=0$
is a characteristic flow, which was called ``point transformations''
by Dirac, in his hamiltonian picture \cite{Dirac}. The generator
of the point transformation \eqref{eq:2.47} is given by $G=\Phi_{z}\epsilon^{z}$,
since $\delta\xi=\left\{ \xi,G\right\} $ reproduces \eqref{eq:2.47}
when $\Phi_{z}=0$.

If the original system has lagrangian symmetries, e.g. local gauge
symmetries in field theories, we expect these symmetries manifest
themselves as canonical symmetries, in this case characteristic flows
in the HJ picture. In this case, eq. \eqref{eq:2.2} for $\delta\tau=0$
yields the Lie equation 
\begin{equation}
\delta L=\frac{\delta L}{\delta q^{i}}\delta q^{i}+\frac{d}{d\tau}\left(\frac{\delta L}{\delta q_{,\left(I+1\right)}^{i}}\delta q_{,I}^{i}\right)=0,\label{eq:2.48}
\end{equation}
and it has to be satisfied by the point transformation \eqref{eq:2.47}
for $\delta q^{i}=\epsilon^{z}X_{z}q^{i}=\left\{ q^{i},G\right\} ^{*}$.

\section{Forms of relativistic dynamics\label{sec:The-null-plane-dynamics}}

\subsection{Poincaré generators in field theories}

In sec. \ref{sec:The-HJ-formalism}, we revised the HJ theory for
singular systems in a classical mechanical background. But since we
propose the study of Podolsky's electrodynamics, our particular interest
rests in second-order relativistic field theories, described by the
functional
\begin{equation}
A\left[\phi\right]\equiv\int_{\Omega}L\left[x^{\mu},\phi^{a}\left(x\right),\phi_{\mu}^{a}\left(x\right),\phi_{\mu\nu}^{a}\left(x\right)\right]d\omega.\label{eq:3.1}
\end{equation}
In \eqref{eq:3.1}, $\Omega$ is a volume of an $\left(d+1\right)$-dimensional
Minkowski space-time $\mathcal{M}$ spanned by a rectangular coordinate
system $x^{\mu}=\left(x^{0},x^{1},x^{2},\cdots,x^{d}\right)$ with
volume element $d\omega\equiv dx^{0}dx^{1}\cdots dx^{d}$. The Lagrangian
density $L$ depends generally on the space-time coordinates $x^{\mu}$,
the fields $\phi^{a}\left(x\right)$ at each point $x\in\Omega$,
its first $\left(\phi_{\mu}^{a}\equiv\partial_{\mu}\phi^{a}\right)$
and second $\left(\phi_{\mu\nu}^{a}\equiv\partial_{\mu}\partial_{\nu}\phi^{a}\right)$
derivatives. If we are dealing with a relativistic field, $L$ must
be Lorentz invariant. The action \eqref{eq:3.1} is distinct from
\eqref{eq:2.1} because the domain is a set of fields of several variables,
and the integral itself is a multiple integral. For this paper, the
metric has signature $\left(+---\right)$.

With \eqref{eq:3.1}, the natural path for a HJ formalism in Carathéodory's
point of view would be to define equivalent actions. But unlike the
case of classical mechanics, there is not a unique way of implementing
Carathéodory's programme in a covariant way. The simplest method,
introduced by Weyl \cite{Weyl}, is by adding to $A$ a simple boundary
term. A second way, developed by Carathéodory himself \cite{Hund},
would be adding a determinant to the Lagrangian density. Each method
leads to distinct definitions of geodesic fields, resulting in distinct
HJ theories. Carathéodory's choice leads to a Hamiltonian function
defined by the determinant of the energy-momentum tensor. Weyl's choice
is the usual one, because it is analog to the classical mechanical
method. Moreover, it does not suffer from important limitations of
the alternative approach, e.g. the requirement of positivity of the
Lagrangian density. But even if we choose Weyl's definition of geodesic
fields, an important characteristic of relativistic theories makes
impracticable the construction of a covariant HJ theory with constraints.
These theories are invariant under any particular choice of global
evolution parameter, or time. In this case, the HJ equation becomes
an identity, not a PDE. An attempt to solve this problem is also done
in \cite{Hund}, but it relies on unphysical assumptions if applied
to fundamental fields. Until now, no satisfactory way of defining
a covariant HJ formalism for constrained systems is at hand.

This problem is closely related to the problem of construction of
covariant Hamiltonian formalisms with constraints. Dirac was the first
to conclude that a consistent Hamiltonian approach for field theories
requires a choice of relativistic dynamics \cite{rel.dyn.}. Such
choice is essentially a choice of \textquotedbl{}time\textquotedbl{}
parameter $\tau$ as a linear combination of the space-time axes,
defined in the direction of a vector field $u\left(x\right)$. This
vector may be related to the world line of a physical observer in
a referential frame, but this is not necessarily the case: a given
frame of reference can use different forms of dynamics to describe
a physical system. Along with the vector $u$, one chooses a family
$\Sigma$ of $d$-dimensional surfaces orthogonal to $u\left(x\right)$
at each point $x\in\Omega$. Each member $\Sigma_{\tau}\in\Sigma$
must be labeled uniquely by a value of $\tau$, then we must choose
the family so that the world line of a given observer intersects each
family member once.

When the vector $u$ and the family $\Sigma$ are chosen, the symmetry
group of the space-time may be decomposed. If $T_{\mu\nu}$ are the
components of the energy-momentum tensor the time evolution generator,
the hamiltonian function, is defined by the double projection
\begin{equation}
H\equiv\int_{\Sigma_{\tau}}d\sigma\thinspace u_{\mu}T_{\thinspace\thinspace\nu}^{\mu}u^{\nu},\label{eq:3.2}
\end{equation}
where $d\sigma_{x}$ is the d-volume element of $\Sigma_{\tau}$.
If $w$ is a unit vector orthogonal to $u$,
\begin{equation}
P_{w}\equiv\int_{\Sigma_{\tau}}d\sigma\thinspace u_{\mu}T_{\thinspace\thinspace\nu}^{\mu}w^{\nu}\label{eq:3.3}
\end{equation}
defines the linear momentum in the direction of $w$, which generates
translations in $\Sigma_{\tau}$ in the $w$ direction.

Let $\left\{ v_{i}\left(x\right)\right\} $ be a base of vector fields
defined in $\Sigma_{\tau}$ for a given point $x$, where $i,j=1,\cdots,d$.
In this case,
\begin{equation}
L_{ij}\equiv\int_{\Sigma_{\tau}}d\sigma\thinspace u_{\mu}\left(T_{\thinspace\thinspace\alpha}^{\mu}x_{\beta}-T_{\thinspace\thinspace\beta}^{\mu}x_{\alpha}\right)v_{i}^{\alpha}v_{j}^{\beta}\label{eq:3.4}
\end{equation}
is the orbital angular momentum matrix, and
\begin{equation}
S_{ij}\equiv-\int_{\Sigma_{\tau}}d\sigma\thinspace u_{\mu}\left[\frac{\delta L}{\delta\phi_{\mu}^{a}}\phi_{\nu}^{a}\left(J_{\alpha\beta}\right)_{\thinspace\thinspace\gamma}^{\nu}x^{\gamma}\right]v_{i}^{\alpha}v_{j}^{\beta}\label{eq:3.5}
\end{equation}
is the spin matrix of the fields, where $\left(J_{\alpha\beta}\right)_{\thinspace\thinspace\gamma}^{\nu}$
are the generators of the Lorentz group. The matrix $M_{ij}=L_{ij}+S_{ij}$
generates rotations in $\Sigma_{\tau}$. We also have boosts, generated
by
\begin{equation}
B_{j}\equiv\int_{\Sigma_{\tau}}d\sigma\thinspace u_{\mu}\left[T_{\thinspace\thinspace\alpha}^{\mu}x_{\beta}-T_{\thinspace\thinspace\beta}^{\mu}x_{\alpha}-\frac{\delta L}{\delta\phi_{\mu}^{a}}\phi_{\nu}^{a}\left(J_{\alpha\beta}\right)_{\thinspace\thinspace\gamma}^{\nu}x^{\gamma}\right]u^{\alpha}v_{j}^{\beta},\label{eq:3.6}
\end{equation}
which are pseudo-rotations in the plane defined by $v_{j}$ and $u$.

\subsection{Null-plane dynamics}

Because of the causal structure of Minkowski's metric, there is not
a unique way of defining a hamiltonian dynamics. The most simple and
used one was called by Dirac \cite{rel.dyn.} the instant-form dynamics.
It is the dynamics of fields in 3-dimensional euclidian spaces orthogonal
to the time axis $t=x^{0}/c$, the time measured by a clock at rest
with respect to the laboratory. In this case, $u$ is given by the
components $\left(1,0,0,0\right)$, while the base vectors $v_{i}$
are the usual $\left\{ x^{1},x^{2},x^{3}\right\} $ axes. The hamiltonian
function is given by $H=\int_{\Sigma_{\tau}}d\sigma\thinspace T_{00}$,
and the linear momenta are just $P_{i}=\int_{\Sigma_{\tau}}d\sigma\thinspace T_{0i}$.
Generators of rotations and boosts are also immediately calculated
from \eqref{eq:3.4}, \eqref{eq:3.5} and \eqref{eq:3.6}.

The null-plane dynamics, on the other hand, is the form of dynamics
where $u$ lies in the light-cone. We define time as the parameter
\begin{equation}
x^{+}=\frac{1}{\sqrt{2}}\left(x^{0}+x^{3}\right),\label{eq:3.7}
\end{equation}
so $u$ takes the form of the axis
\begin{equation}
u^{\mu}=\frac{1}{\sqrt{2}}\left(1,0,0,1\right).\label{eq:3.8}
\end{equation}
The equation $x^{+}=const.$ defines a characteristic hyper-surface
$\Sigma_{x^{+}}$ orthogonal to the $u$ axis. This surface is called
the null-plane, and it represents an electromagnetic wave front in
vacuum.

With this choice of evolution parameter and characteristic surfaces
it is convenient to choose an appropriate coordinate system. Let $x^{\mu}$
be the rectangular coordinates $\left(x^{0},x^{1},x^{2},x^{3}\right)$,
and let us consider the following transformation 
\begin{equation}
y^{\mu}=\Gamma_{\,\,\nu}^{\mu}x^{\nu},\label{eq:3.9}
\end{equation}
with 
\begin{equation}
\Gamma=\frac{1}{\sqrt{2}}\left(\begin{array}{ccc}
1 & 0 & 1\\
1 & 0 & -1\\
0 & \mathbf{I}_{2\times2}\sqrt{2} & 0
\end{array}\right),\label{eq:3.10}
\end{equation}
in which $\mathbf{I}$ is the $2\times2$ identity matrix, and $y^{\mu}$
is defined by the set $\left(x^{+},x^{-},x^{1},x^{2}\right)$. We
see that $x^{\pm}=1/\sqrt{2}\left(x^{0}\pm x^{3}\right)$, while the
remaining coordinates are unchanged. The coordinates $y^{\mu}$ are
called null-plane coordinates.

The null-plane dynamics was mistaken, for some time, with a limiting
process known as the infinite momentum frame \cite{Steinhardt}. But
the transformation \eqref{eq:3.9} is not a reference choice. It is
actually a parameter choice with a combining coordinate system. It
is not even a Lorentz transformation, since the metric in $y$ is
given by 
\begin{equation}
\eta=\left(\begin{array}{ccc}
0 & 1 & 0\\
1 & 0 & 0\\
0 & 0 & -\mathbf{I}_{2\times2}
\end{array}\right).\label{eq:3.11}
\end{equation}
With this metric, the norm of a Lorentz vector is not a quadratic
form in the temporal $\left(+\right)$ and longitudinal $\left(-\right)$
components, but remains quadratic in the transverse components 1 and
2:
\[
A^{2}=\eta_{\mu\nu}A^{\mu}A^{\nu}=2A^{+}A^{-}-\left(A^{i}\right)^{2},\,\,\,\,\,\,\,\,\,\,\,\,\,\,i=1,2.
\]
We stress that the metric \eqref{eq:3.11} is the metric used to raise
and lower indexes in the null-plane coordinates. 

For example, the D'Lambertian operator assumes the form 
\begin{equation}
\square\equiv\partial_{\mu}\partial^{\mu}=\partial_{+}\partial^{+}+\partial_{-}\partial^{-}-\nabla^{2}=2\partial_{+}\partial_{-}-\nabla^{2},\label{eq:3.12}
\end{equation}
where $\nabla\equiv\partial_{i}\partial_{i}$. In this case, the null-plane
dynamics is distinct of the usual instant-form. The Klein-Gordon equation,
for instance, 
\[
\left(\square+m^{2}\right)\phi=0,
\]
has the null-plane form 
\[
\partial_{+}\partial_{-}\phi=\frac{1}{2}\left(\nabla^{2}-m^{2}\right)\phi.
\]
This is a first-order equation in the temporal coordinate $x^{+}$,
therefore the initial value problem is not a Cauchy problem, but a
characteristic value problem. To fix a unique solution, it is necessary
to provide a field configuration in a plane $x^{+}=const.$, and a
second configuration in some $x^{-}=const.$ plane, for example.

In the null-plane we may write expressions for the hamiltonian function
and the linear momenta, based on eqs. \eqref{eq:3.2} and \eqref{eq:3.3}.
The $x^{+}$ evolution generator is given by
\begin{equation}
H=\int_{\Sigma}d\sigma\thinspace T_{\thinspace\thinspace+}^{+}=\int_{\Sigma}d\sigma\thinspace T_{-+},\thinspace\thinspace\thinspace\thinspace\thinspace\thinspace\thinspace\thinspace d\sigma=dx^{-}dx^{1}dx^{2},\label{eq:3.13}
\end{equation}
while translations are generated by
\begin{equation}
P_{-}=\int_{\Sigma}d\sigma\thinspace T_{\thinspace\thinspace-}^{+}=\int_{\Sigma}d\sigma\thinspace T_{--},\thinspace\thinspace\thinspace\thinspace P_{i}=\int_{\Sigma}d\sigma\thinspace T_{\thinspace\thinspace i}^{+}=\int_{\Sigma}d\sigma\thinspace T_{-i}.\label{eq:3.14}
\end{equation}
Generators of rotations and boosts may be calculated with eqs. \eqref{eq:3.4},
\eqref{eq:3.5} and \eqref{eq:3.6}.

\section{Podolsky's electromagnetic theory on the null-plane\label{sec:Podolsky's-electromagnetic-theor}}

\subsection{Conjugated momenta\label{sub:4.1}}

Podolsky's electromagnetic theory is described by the Lagrangian density
\begin{equation}
L=-\frac{1}{4}F_{\mu\nu}F^{\mu\nu}+\frac{1}{2}a^{2}\partial_{\lambda}F^{\mu\lambda}\partial^{\gamma}F_{\mu\gamma},\,\,\,\,\,\,\,\,\,\,\,\,\,\,\,\,\,\,\,F_{\mu\nu}\equiv\partial_{\mu}A_{\nu}-\partial_{\nu}A_{\mu},\label{eq:4.1}
\end{equation}
in which $A_{\mu}$ are the fundamental fields, components of a $U\left(1\right)$
gauge connection. The fields $F_{\mu\nu}$ are components of the curvature
field strength related to the connection field. Therefore, the density
\eqref{eq:4.1} is second-order in the fields $A_{\mu}$, and preserves
the $U\left(1\right)$ gauge symmetry. In fact, all relativistic second-order
Lagrangians with the $U\left(1\right)$ symmetry are equivalent to
\eqref{eq:4.1} \cite{Cuzinatto}.

The action related to \eqref{eq:4.1} is given by 
\begin{equation}
A=\int_{\Omega}d\omega\left(-\frac{1}{4}F_{\mu\nu}F^{\mu\nu}+\frac{1}{2}a^{2}\partial_{\lambda}F^{\mu\lambda}\partial^{\gamma}F_{\mu\gamma}\right),\label{eq:4.2}
\end{equation}
where $\Omega$ is a 4-volume in Minkowski space-time with 4-volume
element $d\omega$. The general first variation of this functional,
analogue to eq. \eqref{eq:2.2}, is given by 
\begin{eqnarray}
\delta A & = & \int_{\Omega}d\omega\frac{\delta L}{\delta A_{\mu}}\left(\delta-\delta x^{\gamma}\partial_{\gamma}\right)A_{\mu}\nonumber \\
 &  & +\int_{\Omega}d\omega\partial_{\gamma}\left[\frac{\delta L}{\delta\left(\partial_{\gamma}A_{\mu}\right)}\delta A_{\mu}+\frac{\delta L}{\delta\left(\partial_{\gamma}\partial_{\nu}A_{\mu}\right)}\delta\left(\partial_{\nu}A_{\mu}\right)-H_{\,\,\mu}^{\gamma}\delta x^{\mu}\right],\label{eq:4.3}
\end{eqnarray}
where 
\begin{equation}
\frac{\delta}{\delta A_{\mu}}=\frac{\partial}{\partial A_{\mu}}-\partial_{\gamma}\frac{\partial}{\partial\left(\partial_{\gamma}A_{\mu}\right)}+\partial_{\gamma}\partial_{\lambda}\frac{\partial}{\partial\left(\partial_{\gamma}\partial_{\lambda}A_{\mu}\right)}\label{eq:4.4}
\end{equation}
is the Lagrange derivative up to the second-order derivative term,
and 
\begin{equation}
H_{\,\,\beta}^{\alpha}\equiv\frac{\delta L}{\delta\left(\partial_{\alpha}A_{\mu}\right)}\partial_{\beta}A_{\mu}+\frac{\delta L}{\delta\left(\partial_{\alpha}\partial_{\gamma}A_{\mu}\right)}\partial_{\beta}\partial_{\gamma}A_{\mu}-L\delta_{\beta}^{\alpha}\label{eq:4.5}
\end{equation}
is the general form of the energy-momentum tensor density.

Observing \eqref{eq:4.5}, we have the following covariant momenta\begin{subequations}\label{eq:4.6}
\begin{gather}
\pi^{\mu\nu}=\frac{\delta\mathrm{L}}{\delta\left(\partial_{\mu}A_{\nu}\right)}=F^{\mu\nu}-2a^{2}\eta^{\omega\gamma}\Upsilon_{\beta\omega}^{\alpha\mu}\Delta_{\alpha\gamma}^{\nu\lambda}\partial_{\lambda}\partial_{\rho}F^{\beta\rho},\label{eq:4.6a}\\
\pi^{\mu\nu\lambda}=\frac{\delta\mathrm{L}}{\delta\left(\partial_{\mu}\partial_{\nu}A_{\lambda}\right)}=2a^{2}\eta^{\omega\gamma}\Upsilon_{\beta\omega}^{\alpha\mu}\Delta_{\alpha\gamma}^{\nu\lambda}\partial_{\rho}F^{\beta\rho},\label{eq:4.6b}
\end{gather}
\end{subequations}where we use the symbols 
\begin{equation}
\Upsilon_{\mu\nu}^{\alpha\beta}\equiv\frac{1}{2}\left(\delta_{\mu}^{\alpha}\delta_{\nu}^{\beta}-\delta_{\mu}^{\beta}\delta_{\nu}^{\alpha}\right),\,\,\,\,\,\,\,\,\,\,\,\Delta_{\mu\nu}^{\alpha\beta}\equiv\frac{1}{2}\left(\delta_{\mu}^{\alpha}\delta_{\nu}^{\beta}+\delta_{\mu}^{\beta}\delta_{\nu}^{\alpha}\right).\label{eq:4.7}
\end{equation}
In this case, we may also write 
\begin{equation}
H_{\alpha\beta}=F_{\,\,\alpha}^{\mu}A_{\mu,\beta}+a^{2}\eta_{\alpha\tau}\Upsilon_{\epsilon\nu}^{\tau\mu}\partial^{\epsilon}\partial_{\lambda}F^{\nu\lambda}A_{\mu,\beta}+2a^{2}\eta^{\gamma\nu}\eta_{\alpha\phi}\Delta_{\mu\nu}^{\lambda\phi}\Upsilon_{\psi\gamma}^{\mu\epsilon}\partial_{\tau}F^{\psi\tau}A_{\epsilon,\lambda\beta}-\eta_{\alpha\beta}L,\label{eq:4.8}
\end{equation}
as the non symmetric energy-momentum tensor density. The field equations
are written by
\begin{equation}
\frac{\delta L}{\delta A_{\mu}}=\left[1+a^{2}\square\right]\partial_{\alpha}F^{\mu\alpha}=0.\label{eq:4.9}
\end{equation}

In the null-plane dynamics, where the time axis is the unit vector
$u^{\alpha}=\left(1,0,0,0\right)$ in null-plane coordinates, the
Hamiltonian function is given by the expression
\begin{equation}
H=\int_{\Sigma}d\sigma H_{\,\,\beta}^{\alpha}u_{\alpha}u^{\beta}=\int_{\Sigma}d\sigma H_{\,\,\,+}^{+}.\label{eq:4.10}
\end{equation}
Again, $\Sigma$ is a 3-surface of constant $x^{+}$, and $d\sigma\equiv dx^{-}dx^{1}dx^{2}$
its respective volume element. The Hamiltonian density takes the form
\begin{eqnarray}
\mathcal{H}_{c} & \equiv & H_{\,\,\,+}^{+}=\pi^{\mu+}A_{\mu,+}+\pi^{\mu+\nu}A_{\mu,+\nu}-L\nonumber \\
 & = & \left[\pi^{\mu+}-\partial_{-}\pi^{\mu+-}-\partial_{i}\pi^{\mu+i}\right]A_{\mu,+}+\pi^{\mu++}A_{\mu,++}-L,\label{eq:4.11}
\end{eqnarray}
where $i=1,2$. This Hamiltonian is precisely of the form $\mathcal{H}_{c}=p^{\mu}\bar{A}_{\mu}+\pi^{\mu}\left(\partial_{+}\bar{A}_{\mu}\right)-L$,
with\begin{subequations}\label{eq:4.12} 
\begin{gather}
p^{\mu}\equiv\pi^{\mu+}-\partial_{-}\pi^{\mu+-}-\partial_{i}\pi^{\mu+i},\\
\pi^{\mu}\equiv\pi^{\mu++}.
\end{gather}
\end{subequations}Then, we identify $\left(p^{\mu},\pi^{\mu}\right)$
as the momenta conjugated to the variables $\left(A_{\mu},\bar{A}_{\mu}\right)$,
respectively. They have the explicit expressions\begin{subequations}\label{eq:4.13}
\begin{gather}
\pi^{\mu}=a^{2}\eta^{\mu+}\partial_{\rho}F^{+\rho},\label{eq:4.13a}\\
p^{\mu}=F^{\mu+}+\partial_{+}\pi^{\mu}+2a^{2}\left(\delta_{\beta}^{\mu}\partial^{+}-\frac{1}{4}\delta_{\beta}^{+}\partial^{\mu}-\frac{1}{2}\eta^{\mu\alpha}\Delta_{\alpha\beta}^{+\lambda}\partial_{\lambda}\right)\partial_{\rho}F^{\beta\rho}.\label{eq:4.13b}
\end{gather}
\end{subequations}

\subsection{The HJ equations\label{sub:The-HJ-equations-1}}

If we write \eqref{eq:4.13} in the null-plane components, we have
the following conjugated momenta:\begin{subequations}\label{eq:4.14}
\begin{gather}
\pi^{+}=0,\\
\pi^{i}=0,\\
\pi^{-}=a^{2}\partial_{\lambda}F^{+\lambda},\\
p^{+}=\partial_{-}\pi^{-},\\
p^{-}=F^{-+}+2a^{2}\partial^{+}\partial_{\lambda}F^{-\lambda},\\
p^{i}=F^{i+}+2a^{2}\partial^{+}\partial_{\lambda}F^{i\lambda}-\partial_{i}\pi^{-}.
\end{gather}
\end{subequations}These are the variables conjugated to $\bar{A}_{+}$,
$\bar{A}_{i}$, $\bar{A}_{-}$, $A_{+}$, $A_{-}$ and $A_{i}$ respectively.
We observe that the momenta $\pi^{+}$, $\pi^{i}$ and $p^{+}$ are
not invertible, thus, we associate them to the canonical constraints\begin{subequations}\label{eq:4.15}
\begin{gather}
\pi^{+}=0,\,\,\,\,\pi^{i}=0,\,\,\,\,\,p^{+}-\partial_{-}\pi^{-}=0.
\end{gather}
\end{subequations}

From $p^{-}$, on the other hand, we obtain an expression for $\partial_{+}\bar{A}_{-}$,
\begin{equation}
\partial_{+}\bar{A}_{-}=\left(2a^{2}\partial_{-}\right)^{-1}\left[p^{-}-\bar{A}_{-}+\left(1-2a^{2}\nabla^{2}\right)\partial_{-}A_{+}+2a^{2}\partial_{-}\partial_{-}\bar{A}_{+}+2a^{2}\partial_{-}\partial_{i}\bar{A}_{i}\right],\label{eq:4.16}
\end{equation}
where $\left(2a^{2}\partial_{-}\right)^{-1}$ represents a Green's
function of the operator $2a^{2}\partial_{-}$, which depends on a
set of initial/boundary conditions, the characteristic data of the
fields. Without specifying these data, \eqref{eq:4.16} is actually
a family of solutions. We will need to fix the characteristic data
later when analysing the HJ equations of the system, and in fact the
expression \eqref{eq:4.16} is not required in any of the following
analysis. In this case, we let the initial/boundary analysis for later.

Now, the expression for $p^{i}$ can be written in the form 
\begin{equation}
p^{i}=F_{-i}+\partial_{i}\pi^{-}+2a^{2}\partial_{-}\left(2\partial_{-}\bar{A}_{i}-\partial_{i}\bar{A}_{-}-\partial_{-}\partial_{i}\bar{A}_{+}+\partial_{j}F_{ij}\right).\label{eq:4.17}
\end{equation}
This expression does not provide any velocity, therefore, \eqref{eq:4.15}
and \eqref{eq:4.17} constitute the following set of constraints\begin{subequations}\label{eq:4.18}
\begin{gather}
\phi_{1}=\pi^{+},\\
\phi_{2}^{i}=\pi^{i},\\
\phi_{3}=p^{+}-\partial_{-}\pi^{-},\\
\phi_{4}^{i}=p^{i}-\partial_{i}\pi^{-}+F_{i-}+2a^{2}\partial_{-}\left[\partial_{i}\bar{A}_{-}-2\partial_{-}\bar{A}_{i}+\partial_{i}\partial_{-}A_{+}-\partial_{j}F_{ij}\right].
\end{gather}
\end{subequations}Since
\begin{equation}
\pi^{\mu}\equiv\frac{\partial S}{\partial\bar{A}_{\mu}},\,\,\,\,p^{\mu}\equiv\frac{\partial S}{\partial A_{\mu}},\label{eq:4.19}
\end{equation}
eqs. \eqref{eq:4.18} are a set of HJ equations.

Another HJ equation is necessary, the one related to the canonical
Hamiltonian density \eqref{eq:4.11}, which has the explicit form
\begin{equation}
\mathcal{H}_{c}=p^{\mu}\bar{A}_{\mu}+\pi^{-}\left(\partial_{-}\bar{A}_{+}+\partial_{i}\bar{A}_{i}-\nabla^{2}A_{+}\right)-\frac{1}{2}a^{2}\partial_{\lambda}F^{i\lambda}\partial^{\gamma}F_{i\gamma}+\frac{1}{4}F_{\mu\nu}F^{\mu\nu},\label{eq:4.20}
\end{equation}
where $\nabla^{2}=\partial_{i}\partial_{i}$. Note that this density
has no explicit dependency of any velocity, the same happening with
the densities in \eqref{eq:4.18}. Then, there is no need for \eqref{eq:4.16}
in the construction of the constraints. In this case, we simply write
the set of HJ equations\begin{subequations}\label{eq:4.21}
\begin{gather}
\phi_{0}=p_{0}+\mathcal{H}_{c},\\
\phi_{1}=\pi^{+},\\
\phi_{2}^{i}=\pi^{i},\\
\phi_{3}=p^{+}-\partial_{-}\pi^{-},\\
\phi_{4}^{i}=p^{i}-\partial_{i}\pi^{-}+F_{i-}+2a^{2}\partial_{-}\left[\partial_{i}\bar{A}_{-}-2\partial_{-}\bar{A}_{i}+\partial_{-}\partial_{i}A_{+}-\partial_{j}F_{ij}\right],
\end{gather}
\end{subequations}where $p_{0}\equiv\partial S/\partial x^{+}$ is
the momentum conjugated to the time variable $x^{+}$.

\subsection{Generalised Brackets\label{sub:Generalized-Brackets}}

To be a complete integrable system, \eqref{eq:4.21} must be a complete
involutive set with the complete Poisson brackets
\begin{eqnarray}
\left\{ F,G\right\}  & \equiv & \int_{\Sigma}d\sigma_{x}\left[\frac{\partial F}{\partial A_{\mu}\left(x\right)}\frac{\partial G}{\partial p^{\mu}\left(x\right)}-\frac{\partial G}{\partial A_{\mu}\left(x\right)}\frac{\partial F}{\partial p^{\mu}\left(x\right)}+\right.\nonumber \\
 &  & \left.+\frac{\partial F}{\partial\bar{A}_{\mu}\left(x\right)}\frac{\partial G}{\partial\pi^{\mu}\left(x\right)}-\frac{\partial G}{\partial\bar{A}_{\mu}\left(x\right)}\frac{\partial F}{\partial\pi^{\mu}\left(x\right)}\right],\label{eq:4.22}
\end{eqnarray}
for any observables $F$ and $G$ of the complete phase space of the
variables $\left(A_{\mu},\bar{A}_{\mu},p^{\mu},\pi^{\mu}\right)$.
Clearly, the fundamental PB are
\begin{equation}
\left\{ A_{\mu}\left(x\right),p^{\nu}\left(y\right)\right\} =\left\{ \bar{A}_{\mu}\left(x\right),\pi^{\nu}\left(y\right)\right\} =\delta_{\mu}^{\nu}\delta^{3}\left(x-y\right),\label{eq:4.23}
\end{equation}
in which $\delta^{3}\left(x-y\right)\equiv\delta\left(x^{-}-y^{-}\right)\delta\left(x^{1}-y^{1}\right)\delta\left(x^{2}-y^{2}\right)$
is the appropriate Dirac's delta.

When calculating the matrix $M\equiv\left\{ \phi_{A},\phi_{B}\right\} $,
with $\phi_{A}=\left(\phi_{0},\phi_{1},\phi_{2}^{i},\phi_{3},\phi_{4}^{i}\right)$,
we see that $\phi_{1}$ is in involution, but not the remaining constraints.
Particularly, the subset $\left(\phi_{2}^{i},\phi_{3},\phi_{4}^{i}\right)$
is non involutive, but it gives rise to a singular sub-matrix, indicating
that a particular linear combination of these constraints is integrable.
However, we have the subset $\left(\phi_{2}^{i},\phi_{4}^{i}\right)$,
which obeys the relations\begin{subequations}\label{eq:4.24}
\begin{gather}
\left\{ \phi_{2}^{i}\left(x\right),\phi_{4}^{j}\left(y\right)\right\} =-4a^{2}\eta^{ij}\partial_{-}^{x}\partial_{-}^{x}\delta^{3}(x-y),\\
\left\{ \phi_{4}^{i}\left(x\right),\phi_{4}^{j}\left(y\right)\right\} =2\eta^{ij}\left[1-2a^{2}\nabla_{x}^{2}\right]\partial_{-}^{x}\delta^{3}\left(x-y\right).
\end{gather}
\end{subequations}This particular subset of HJ equations give rise
to the matrix
\begin{equation}
M_{IJ}^{ij}\equiv2\eta^{ij}\left(\begin{array}{cc}
0 & -2a^{2}\partial_{-}^{x}\partial_{-}^{x}\\
2a^{2}\partial_{-}^{x}\partial_{-}^{x} & \partial_{-}^{x}\left[1-2a^{2}\nabla_{x}^{2}\right]
\end{array}\right)\delta^{3}\left(x-y\right),\,\,\,\,\,\,I,J=2,4.\label{eq:4.25}
\end{equation}

A matrix $\left(M^{-1}\right)_{ij}^{IJ}$ obeying
\begin{equation}
\int d^{3}zM_{IJ}^{ij}\left(x,z\right)\left(M^{-1}\right)_{jk}^{JK}\left(z,y\right)=\int d^{3}z\left(M^{-1}\right)_{jk}^{JK}\left(x,z\right)M_{IJ}^{ij}\left(z,y\right)=\delta_{k}^{i}\delta_{I}^{K}\delta^{3}\left(x-y\right),\label{eq:4.26}
\end{equation}
if unique, is defined as the inverse matrix of $M_{IJ}^{ij}$. A solution
for \eqref{eq:4.26} is given by
\begin{equation}
\left(M^{-1}\right)_{ij}^{IJ}\left(x,y\right)\equiv\frac{1}{2}\eta_{ij}\left(\begin{array}{cc}
\alpha\left(x,y\right) & \beta\left(x,y\right)\\
\gamma\left(x,y\right) & 0
\end{array}\right),\label{eq:4.27}
\end{equation}
with\begin{subequations}\label{eq:4.28}
\begin{gather}
\alpha\left(x,y\right)=\frac{1}{4a^{2}}\left|x^{-}-y^{-}\right|^{2}\epsilon\left(x^{-}-y^{-}\right)\left[1-2a^{2}\nabla_{x}^{2}\right]\delta^{2}\left(\mathbf{x}-\mathbf{y}\right),\\
\beta\left(x,y\right)=-\gamma\left(x,y\right)=\frac{1}{a^{2}}\left|x^{-}-y^{-}\right|\delta^{2}\left(\mathbf{x}-\mathbf{y}\right).
\end{gather}
\end{subequations}In \eqref{eq:4.28} we have $\delta^{2}\left(\mathbf{x}-\mathbf{y}\right)\equiv\delta\left(x^{1}-y^{1}\right)\delta\left(x^{2}-y^{2}\right)$,
and the sign function
\begin{equation}
\epsilon\left(x-y\right)\equiv\left\{ \begin{array}{l}
1\,\,\,\,\,\,\,\,\,\,\,\,\,\,\,\,\,\,x>y\\
0\,\,\,\,\,\,\,\,\,\,\,\,\,\,\,\,\,\,x=y\\
-1\,\,\,\,\,\,\,\,\,\,\,\,\,\,x<y
\end{array}\right..\label{eq:4.29}
\end{equation}

To each of the functions \eqref{eq:4.28} we may add arbitrary functions
of $x^{+}$, $x^{1}$, and $x^{2}$, and polynomials of $x^{-}$ up
to second-order. Therefore, we do not have a unique inverse. These
functions can be fixed using proper boundary conditions for the fields
in a null plane of constant $x^{-}$. We choose these conditions to
be $\partial_{-}A_{\mu}=\partial_{-}^{2}A_{\mu}=\partial_{-}^{3}A_{\mu}=0$
for $x^{-}\rightarrow-\infty$. In this case, we may treat \eqref{eq:4.27}
as an appropriate inverse matrix.

The GB related to the subset $\left(\phi_{2}^{i},\phi_{4}^{i}\right)$
are given by
\begin{equation}
\left\{ F,G\right\} ^{*}\equiv\left\{ F,G\right\} -\iint_{\Sigma}d\sigma_{x}d\sigma_{y}\left\{ F,\phi_{I}^{i}\left(x\right)\right\} \left(M^{-1}\right)_{ij}^{IJ}\left(x,y\right)\left\{ \phi_{J}^{j}\left(y\right),G\right\} ,\label{eq:4.30}
\end{equation}
and result in the fundamental relations\begin{subequations}\label{eq:4.31}
\begin{eqnarray}
\left\{ A_{\mu}\left(x\right),\bar{A}_{\nu}\left(y\right)\right\} ^{\ast} & = & -\frac{1}{2}\eta_{\mu j}\delta_{\nu}^{j}\beta\left(x,y\right),\\
\left\{ A_{\mu}\left(x\right),p^{\nu}\left(y\right)\right\} ^{\ast} & = & \delta_{\mu}^{\nu}\delta^{3}\left(x-y\right),\\
\left\{ \bar{A}_{\mu}\left(x\right),\bar{A}_{\nu}\left(y\right)\right\} ^{\ast} & = & \frac{1}{2}\eta_{\mu j}\delta_{\nu}^{j}\alpha\left(x,y\right)+\Delta_{\mu\nu}^{j-}\partial_{j}^{x}\beta\left(x,y\right),\\
\left\{ \bar{A}_{\mu}\left(x\right),p^{\nu}\left(y\right)\right\} ^{\ast} & = & \frac{1}{2}\delta_{\mu}^{j}\delta_{-}^{\nu}\partial_{i}^{x}\beta\left(x,y\right)\\
 &  & -a^{2}\delta_{\mu}^{i}\left[\delta_{k}^{\nu}\partial_{k}^{x}\partial_{i}^{x}-\delta_{+}^{\nu}\partial_{i}^{x}\partial_{-}^{x}+\delta_{i}^{\nu}\left(\frac{1}{2a^{2}}-\nabla_{x}^{2}\right)\right]\partial_{-}^{x}\beta\left(x,y\right),\\
\left\{ \bar{A}_{\mu}\left(x\right),\pi^{\nu}\left(y\right)\right\} ^{\ast} & = & \delta_{\mu}^{\nu}\delta^{3}\left(x-y\right)+a^{2}\delta_{\mu}^{i}\left[\delta_{-}^{\nu}\partial_{i}^{x}-2\delta_{i}^{\nu}\partial_{-}^{x}\right]\partial_{-}^{x}\beta\left(x,y\right).
\end{eqnarray}
\end{subequations}

Since the subset $\left(\phi_{2}^{i},\phi_{4}^{i}\right)$ is eliminated,
we have $\left\{ \phi_{1},\phi_{3}\right\} ^{*}=0$ and $\left\{ \phi_{0},\phi_{1}\right\} ^{*}=\phi_{3}$.
However,
\begin{equation}
\left\{ \phi_{0}\left(x\right),\phi_{3}\left(y\right)\right\} ^{\ast}=-\left[\partial_{-}^{x}p^{-}\left(x\right)+\partial_{i}^{x}p^{i}\left(x\right)\right]\delta^{3}\left(x-y\right).\label{eq:4.32}
\end{equation}
Eq. \eqref{eq:4.32} implies a new HJ equation, named
\begin{equation}
\Phi\left(x\right)\equiv-\partial_{-}^{x}p^{-}\left(x\right)-\partial_{i}^{x}p^{i}\left(x\right)=0.\label{eq:4.33}
\end{equation}
Now we have the extended, and renamed set of constraints\begin{subequations}\label{eq:4.34}
\begin{gather}
\Phi_{0}\equiv p_{0}+\mathcal{H}_{c},\\
\Phi_{1}\equiv\pi^{+}\left(x\right),\\
\Phi_{2}\equiv p^{+}\left(x\right)-\partial_{-}^{x}\pi^{-}\left(x\right),\\
\Phi_{3}\equiv-\partial_{-}^{x}p^{-}\left(x\right)-\partial_{i}^{x}p^{i}\left(x\right).
\end{gather}
\end{subequations}Calculating the GB between them, we see that this
set is complete and in involution with the GB operation. Therefore,
we have completed the task at hand, which was to find a complete integrable
set of HJ equations for the system.

\subsection{The field equations\label{sub:The-field-equations}}

With the GB \eqref{eq:4.30}, the fundamental differential
\begin{equation}
dF\left(x\right)=\int_{\Sigma}d\sigma_{y}\left\{ F\left(x\right),\Phi_{A}\left(y\right)\right\} ^{*}dt^{A}\left(y\right),\,\,\,\,\,\,\,\,\,A=\left(0,1,2,3\right),\label{eq:4.35}
\end{equation}
gives the dynamical evolution of any observable $F\left(A,\bar{A},p,\pi\right)$
of the phase space of the system. The set of HJ equations \eqref{eq:4.34},
represented here by the set $\Phi_{A}\left(A,\bar{A},p,\pi\right)=0$,
is completely integrable due to the analysis made in sec. \ref{sub:Generalized-Brackets},
and provides a set of dynamical generators for the evolution of the
system. They may also be called Hamiltonians of the system. Each Hamiltonian
$\Phi_{A}$ is related to an independent variable (or parameter) $t^{A}$,
which is completely arbitrary in principle. We know, of course, that
some of the parameters must be related to some of the phase space
variables, and, particularly, $t^{0}=x^{+}$ is found when $F=x^{+}$
in \eqref{eq:4.35}. The parameter $t^{3}$, on the other hand, is
not part of the original set of independent variables, since $\Phi_{3}$
is a generator found later in the integrability analysis, and no phase
space variable can be related to it, in principle. Nevertheless, the
parameter space is expanded to contain as many parameters as the number
of involutive constraints of the system, as long as integrability
is assured. Moreover, the Frobenius' theorem implies linear independence
of the independent variables, so the evolution in a given \textquotedbl{}direction\textquotedbl{}
does not depend on other \textquotedbl{}directions\textquotedbl{}
of the parameter space.

For $A_{\mu}$, we have the characteristic equations
\begin{equation}
dA_{\mu}=\bar{A}_{\mu}dt^{0}+\delta_{\mu}^{+}dt^{2}+\left(\delta_{\mu}^{-}\partial_{-}+\delta_{\mu}^{i}\partial_{i}\right)dt^{3}.\label{eq:4.36}
\end{equation}
Since $t^{0}=x^{+}$, time evolution is given by
\begin{equation}
\partial_{+}A_{+}=\bar{A}_{+},\,\,\,\,\,\,\,\,\partial_{+}A_{-}=\bar{A}_{-},\,\,\,\,\,\,\,\,\partial_{+}A_{i}=\bar{A}_{i},\label{eq:4.37}
\end{equation}
as expected. The characteristic equation associated to $\bar{A}_{\mu}$
is
\begin{equation}
d\bar{A}_{\mu}=\delta_{\mu}^{-}\left[\partial_{-}\bar{A}_{+}+\partial_{i}\bar{A}_{i}-\nabla^{2}A_{+}\right]dt^{0}+\delta_{\mu}^{+}dt^{1}+\left[\delta_{\mu}^{-}\partial_{-}+\delta_{\mu}^{i}\partial_{i}\right]dt^{2}.\label{eq:4.38}
\end{equation}
Time evolution alone yields
\begin{equation}
\partial_{+}\bar{A}_{+}=0,\,\,\,\,\partial_{+}\bar{A}_{i}=0,\,\,\,\,\partial_{+}\bar{A}_{-}=\partial_{-}\bar{A}_{+}+\partial_{i}\bar{A}_{i}-\nabla^{2}A_{+},\label{eq:4.39}
\end{equation}
which are expected. Particularly, the $\partial_{+}\bar{A}_{-}$ equation
agrees with \eqref{eq:4.16} when the constraints are used.

The characteristic equations associated to the canonical momenta $p^{\mu}$
and $\pi^{\mu}$ are obtained from \eqref{eq:4.35}, resulting in
the expressions\begin{subequations}\label{eq:4.40}
\begin{gather}
\partial_{+}p^{+}=\partial_{\lambda}F^{\lambda+}-a^{2}\partial_{i}\partial_{-}\partial_{\lambda}F^{\lambda i}-a^{2}\nabla^{2}\partial_{\lambda}F^{\lambda+},\\
\partial_{+}p^{i}=\partial_{-}F^{-i}+\left(1-a^{2}\square\right)\partial_{j}F^{ij},\\
\partial_{+}p^{-}=\partial_{i}F^{i-}.
\end{gather}
\end{subequations}Similarly, for $\pi^{\mu}$ we derive\begin{subequations}\label{eq:4.41}
\begin{gather}
\partial_{+}\pi^{+}=\phi_{3},\\
\partial_{+}\pi^{+}=F_{+-}-p^{-}+a^{2}\partial_{i}\partial_{\lambda}F^{\lambda i},\\
\partial_{+}\pi^{i}=-\phi_{4}^{i}.
\end{gather}
\end{subequations}The relations associated to $\pi^{+}$ and $\pi^{-}$
indicate what we saw from the integrability conditions, i.e., $\phi_{3}$
and $\phi_{4}^{i}$ result from the conditions $d\phi_{1}=0$ and
$d\phi_{2}^{i}=0$. Using the fact that $\partial_{+}\pi^{+}=0$,
$\partial_{+}\pi^{i}=0$ and $\pi^{-}=a^{2}\partial_{\lambda}F^{+\lambda}$,
we obtain\begin{subequations}\label{eq:4.42}
\begin{gather}
p^{+}=a^{2}\partial_{-}\partial_{\lambda}F^{+\lambda},\\
p^{-}=F^{-+}+a^{2}\partial_{\lambda}F^{-\lambda},\\
p^{i}=F^{+i}-a^{2}\partial_{i}\partial_{+}\partial_{\lambda}F^{\lambda+}-2a^{2}\partial_{-}\partial_{+}\partial_{\lambda}F^{\lambda i}.
\end{gather}
\end{subequations}Replacing \eqref{eq:4.42} in \eqref{eq:4.41},
we deduce the field equations 
\begin{equation}
\left(1+a^{2}\square\right)\partial_{\lambda}F^{\mu\lambda}=0.\label{eq:4.43}
\end{equation}
Therefore, the time evolution sector of the characteristic equations
are equivalent to the Lagrangian field equations.

\subsection{Symmetries\label{sub:Symmetries}}

Let us go back to the complete set of dynamical generators \eqref{eq:4.34},
with the canonical Hamiltonian density \eqref{eq:4.20}. The Hamiltonians
$\Phi_{A}$, as shown in \cite{ICHJF}, are generators of canonical
transformations in the complete phase space. The $x^{+}$ \textquotedbl{}time\textquotedbl{}
evolution studied above is a particular case. As usual in constraint
analysis, we are also interested in the canonical transformations
which represent Lagrangian gauge transformations. In this case, we
consider the following infinitesimal transformations
\begin{equation}
\delta F\left(x\right)=\int_{\Sigma}d\sigma_{y}\left\{ F\left(x\right),\Phi_{a}\left(y\right)\right\} ^{*}\delta\omega^{a}\left(y\right),\,\,\,\,\,\,\,\,\,a=\left(1,2,3\right),\label{eq:4.44}
\end{equation}
which are taken with $\delta x^{+}=0$. The generator is given by
\begin{equation}
G\equiv\int_{\Sigma}d\sigma_{x}\Phi_{a}\left(x\right)\delta\omega^{a}\left(x\right),\,\,\,\,\,\,\,\,\,\,a=1,2,3.\label{eq:4.45}
\end{equation}
with $\Phi_{a}=0$.

Now we calculate a fixed point transformation on the Lagrangian density
\eqref{eq:4.1}, given by
\begin{equation}
\delta L=\delta A_{\mu}\frac{\delta L}{\delta A_{\mu}}=\delta A_{\mu}\left(\square+m^{2}\right)\partial_{\alpha}F^{\mu\alpha},\thinspace\thinspace\thinspace\thinspace\thinspace\thinspace\thinspace\thinspace\thinspace\thinspace\thinspace m^{2}\equiv1/a^{2},\label{eq:4.46}
\end{equation}
apart of a total divergence. Remember that $\delta L/\delta A_{\mu}$
is the Lagrange derivative \eqref{eq:4.4}. On the other hand, $\delta A_{\mu}$
is generated by $G$ \emph{via} the generalised brackets, which gives
the result 
\begin{equation}
\delta A_{\mu}=\left\{ A_{\mu},G\right\} ^{*}=\delta_{\mu}^{+}\delta\omega^{2}-\left(\delta_{\mu}^{-}\partial_{-}+\delta_{\mu}^{i}\partial_{i}\right)\delta\omega^{3}.\label{eq:4.47}
\end{equation}

Apart of divergence terms, substitution of \eqref{eq:4.47} in \eqref{eq:4.46}
yields
\[
\delta L=\delta\omega^{2}\left(\square+m^{2}\right)\partial_{\alpha}F^{+\alpha}-\delta\omega^{3}\left(\square+m^{2}\right)\left(\partial_{-}\partial_{\alpha}F^{-\alpha}+\partial_{i}\partial_{\alpha}F^{i\alpha}\right).
\]
Note that the identity $\partial_{\alpha}\partial_{\beta}F^{\alpha\beta}=0$
results in
\[
\partial_{-}\partial_{\alpha}F^{-\alpha}+\partial_{i}\partial_{\alpha}F^{i\alpha}=-\partial_{+}\partial_{\alpha}F^{+\alpha},
\]
so, after some calculation,
\begin{equation}
\delta L=\left(\delta\omega^{2}+\partial_{+}\delta\omega^{3}\right)\left(\square+m^{2}\right)\partial_{\alpha}F^{+\alpha}.\label{eq:4.48}
\end{equation}

The system is gauge invariant under the transformation \eqref{eq:4.47}
if $\delta\mathcal{L}$ is zero in $\Omega$. Of course, the field
equations themselves lead to $\delta\mathcal{L}=0$, but if the symmetry
is understood to be valid outside the solutions of the variational
problem, we must have
\begin{equation}
\delta\omega^{2}=-\partial_{+}\delta\omega^{3}.\label{eq:4.49}
\end{equation}
Supposing $\Lambda\equiv\delta\omega^{3}$ the gauge parameter, we
have
\begin{equation}
\delta\omega^{2}=-\partial_{+}\Lambda,\label{eq:4.50}
\end{equation}
so \eqref{eq:4.47} gives
\begin{equation}
\delta A_{\mu}=-\delta_{\mu}^{+}\partial_{+}\Lambda-\delta_{\mu}^{-}\partial_{-}\Lambda-\delta_{\mu}^{i}\partial_{i}\Lambda=-\partial_{\mu}\Lambda.\label{eq:4.51}
\end{equation}
Therefore, we recover the correct gauge transformations $A_{\mu}\rightarrow A_{\mu}-\partial_{\mu}\Lambda$.

On the other hand, taking $\partial_{+}\Lambda\equiv\dot{\Lambda}$,

\[
\delta\bar{A}_{\mu}=\delta_{\mu}^{+}\delta\omega^{1}-\delta_{\mu}^{-}\partial_{-}\dot{\Lambda}-\delta_{\mu}^{i}\partial_{i}\dot{\Lambda},
\]
and $\delta\omega^{1}$ is still arbitrary. However, if the gauge
transformation for $A_{\mu}$ is given by $\delta A_{\mu}=-\partial_{\mu}\Lambda$,
the same transformation for $\bar{A}_{\mu}$ should be $\delta\bar{A}_{\mu}=-\partial_{\mu}\dot{\Lambda}$
for the sake of consistency. Therefore, we should fix $\delta\omega^{1}=-\ddot{\Lambda}$,
so
\begin{equation}
\delta\bar{A}_{\mu}=-\delta_{\mu}^{+}\ddot{\Lambda}-\delta_{\mu}^{-}\partial_{-}\dot{\Lambda}-\delta_{\mu}^{i}\partial_{i}\dot{\Lambda}=-\partial_{\mu}\dot{\Lambda}.\label{eq:4.52}
\end{equation}
After some calculation, the generator of gauge transformations \eqref{eq:4.45}
becomes
\begin{equation}
G=\int_{\Sigma}d\sigma_{x}\left(\partial_{\mu}\pi^{\mu}\dot{\Lambda}+\partial_{\mu}p^{\mu}\Lambda\right).\label{eq:4.53}
\end{equation}
In fact, the generator \eqref{eq:4.53} reproduces the relations \eqref{eq:4.51}
and \eqref{eq:4.52}.

\section{Final remarks\label{sec:Final-remarks}}

In this paper, we analysed the null-plane canonical structure of Podolsky's
generalised electrodynamics \emph{via} the Hamilton-Jacobi formalism.
The essence of the HJ approach is to understand canonical constraints
as generators of transformations on the phase space. Each generator
is a hamiltonian function responsible to the evolution of the system
along a respective evolution parameter, as the canonic hamiltonian
function which is the generator of time evolution. The flows generated
by the hamiltonians are solutions of the characteristic equations
taken from the fundamental differential \eqref{eq:2.44}. If complete
integrability is assured by application of Frobenius' conditions,
these generators are in involution with the generalised brackets,
closing themselves a Lie algebra. A complete set of involutive constraints
generates transformations that preserve the symplectic structure of
the phase space, therefore becoming canonical transformations, whose
flows are called characteristic flows. In this case, we understand
the HJ formalism as the search for a complete set of involutive hamiltonians
of a singular system.

Applied to Podolsky's theory, the HJ approach starts with the definition
of the constraints \eqref{eq:4.21}. Among these constraints, we identify
two sets of non-involutive ones, leading to the GB \eqref{eq:4.30},
with fundamental relations calculated in \eqref{eq:4.31}. These brackets
are defined only with the application of proper boundary conditions,
$\partial_{-}A_{\mu}=\partial_{-}^{2}A_{\mu}=\partial_{-}^{3}A_{\mu}=0$
for $x^{-}\rightarrow-\infty$. The GB eliminate the set $\left(\phi_{2}^{i},\phi_{4}^{i}\right)$,
but reveals a new constraint given by \eqref{eq:4.33}. The complete
set of hamiltonians, written in \eqref{eq:4.34}, is in involution
with the GB, therefore completing the task of finding the complete
set of generators of the system.

In section \ref{sub:The-field-equations} the canonical field equations
are calculated. It is shown that these characteristic equations, in
the temporal sector, are equal to the Euler-Lagrange equations of
Podolsky's lagrangian.

On the other hand, the evolution of the system in the direction of
the remaining independent variables $\omega^{a}$, with $\delta x^{+}=0$,
is analysed in section \ref{sub:Symmetries}. In this case, the hamiltonians
$\Phi_{a}$ generate characteristic flows defined by \eqref{eq:4.45},
and these flows are symmetries of the system. Since $\Phi_{a}$ form
a complete set of compatible observables in the reduced phase space,
we expect that these transformations are related to the gauge transformations
defined by an invariant field strength $F_{\mu\nu}$. In fact, $\delta L=0$
leads directly to the relation \eqref{eq:4.49} between the former
independent variables. Choosing $\Lambda\equiv\delta\omega^{3}$ as
the gauge parameter, we see that $\omega^{2}$ must have the form
$\delta\omega^{2}=-\dot{\Lambda}$, while $\omega^{1}$ must obey
$\delta\omega^{1}=-\ddot{\Lambda}$. The correct gauge transformations
\eqref{eq:4.51} and \eqref{eq:4.52} are generated by the generating
function \eqref{eq:4.53}.

Here, we stress the fact that the application of the HJ theory not
only provides straightforward results for the complete set of generators
and consistent canonical field equations, but also resulted in the
correct generators of the gauge transformations, \emph{via} the correct
relation among the independent variables of the theory. This last
result was always of some controversy in the literature, since Dirac's
method incorporate an unproven assumption, Dirac's conjecture, which
is the statement that all first-class constraints (correspondent to
our involutive constraints) must contribute to the construction of
the generator. Moreover, no ad-hoc method, as Castellani's procedure
\cite{Castellani}, was necessary. In fact, it seems that Dirac's
conjecture is actually not a defined problem in the HJ formalism.
It is our intent, however, to study the relation between Frobenius'
theorem and this conjecture in the near future.

\section*{Acknowledgements}

M.C.B. thanks CAPES and FAPESP for partial support. B.M.P. thanks
CNPq for partial support. C.E.V. was supported by CNPq process 150407/2016\textminus 5.
G.E.R.G. was supported by VIPRI-UDENAR.


\begin{thebibliography}{10}
\bibitem{Caratheodory}C. Carathéodory. Calculus of Variations and
Partial Differential Equations of the First Order. AMS Chelsea Publishing
2000.

\bibitem{Hund}H. Rund, The Hamilton-Jacobi theory in the calculus
of variations; its role in mathematics and physics, Van Nostrand 1966.

\bibitem{Dirac}P. A. M. Dirac, Generalized hamiltonian dynamics,
\href{http://dx.doi.org/10.4153/CJM-1950-012-1}{Can. J. Math.} 2,
129 (1950) .\\
P. A. M. Dirac, The hamiltonian form of field dynamics, \href{http://dx.doi.org/10.4153/CJM-1951-001-2}{Can. J. Math.}
3, 1 (1951).\\
P. A. M. Dirac, Lectures on Quantum Mechanics, Yeshiva University,
New York (1964).

\bibitem{Bergmann}J. L. Anderson, P. G. Bergmann, Constraints in
Covariant Field Theories, \href{http://dx.doi.org/10.1103/PhysRev.83.1018}{Phys. Rev.}
83, 1018 (1951).\\
P. G. Bergmann, I. Goldberg, Dirac Bracket Transformations in Phase
Space, \href{http://dx.doi.org/10.1103/PhysRev.98.531}{Phys. Rev.}
98, 531 (1955).

\bibitem{Guler-HJ}Y. Güler, Integration of singular systems, \href{http://dx.doi.org/10.1007/BF02727199}{Il Nuovo Cimento B}
107, 1143 (1992).\\
Y. Güler, Canonical formulation of singular systems, \href{http://dx.doi.org/10.1007/BF02722849}{Il Nuovo Cimento B}
107, 1389 (1992).

\bibitem{NICS}M. C. Bertin, B. M. Pimentel, C. E. Valcárcel, Non-involutive
constrained systems and Hamilton-Jacobi formalism, \href{http://dx.doi.org/10.1016/j.aop.2008.09.002}{Ann. Phys.}
323, 3137 (2008).

\bibitem{ICHJF}M. C. Bertin, B. M. Pimentel, C. E. Valcárcel, Involutive
constrained systems and Hamilton-Jacobi formalism, \href{http://dx.doi.org/10.1063/1.4900921}{J. Math. Phys.}
55, 112901 (2014).

\bibitem{general HJ}B. M. Pimentel, R. G. Teixeira, Hamilton-Jacobi
formulation for singular systems with second order lagrangians, \href{http://dx.doi.org/10.1007/BF02749015}{Il Nuovo Cimento B}
111, 841 (1996) .\\
B. M. Pimentel, R. G. Teixeira, Generalization of the Hamilton-Jacobi
approach for higher order singular systems, \href{http://en.sif.it/journals/ncb/econtents/1998/113/06/article/1}{Il Nuovo Cimento B}
113, 805 (1998).\\
M. C. Bertin, B. M. Pimentel, P. J. Pompeia, Hamilton-Jacobi approach
for first order actions and theories with higher derivatives, \href{http://dx.doi.org/10.1016/j.aop.2007.11.003}{Ann. Phys.}
323, 527 (2008).

\bibitem{Ostrogradski}M. Ostrogradski, Mem. Ac. St. Petersbourg 1,
385 (1850). 

\bibitem{Cuzinatto}R. R. Cuzinatto, C. A. M. de Melo, and P. J. Pompeia,
Second order gauge theory, \href{http://dx.doi.org/10.1016/j.aop.2006.07.006}{Ann. Phys.}
322, 1211 (2007).\\
R. R. Cuzinatto, C. A. M. de Melo, L. G. Medeiros, and P. J. Pompeia,
Gauge formulation for higher order gravity, \href{http://dx.doi.org/10.1140/epjc/s10052-007-0441-1}{Eur. Phys. J. C}
53, 99 (2008).

\bibitem{Stelle - Querella}K. S. Stelle, Renormalization of higher-derivative
quantum gravity, \href{http://dx.doi.org/10.1103/PhysRevD.16.953}{Phys. Rev. D}
16, 953 (1977).\\
K. S. Stelle, Classical gravity with higher derivatives, \href{http://dx.doi.org/10.1007/BF00760427}{Gen. Rel. Grav.}
9, 353 (1978).\\
L. Querella, Variational principles and cosmological models in higher-order
gravity, Doctoral Thesis, \href{http://arxiv.org/abs/gr-qc/9902044}{e-print arXiv: gr-qc/9902044},
and references therein.

\bibitem{NMG}E. A. Bergshoe, O. Hohm, and P. K. Townsend, Massive
Gravity in Three Dimensions, \href{http://dx.doi.org/10.1103/PhysRevLett.102.201301}{Phys. Rev. Lett.}
102, 201301 (2009).\\
E. A. Bergshoe, O. Hohm, and P. K. Townsend, More on massive 3D gravity,
\href{http://dx.doi.org/10.1103/PhysRevD.79.124042}{Phys. Rev. D}
79, 124042 (2009).

\bibitem{Bopp}F. Bopp, Eine Lineare Theorie des Elektrons. \href{http://onlinelibrary.wiley.com/doi/10.1002/andp.19404300504/full}{Ann. Phys.}
430, 345 (1940).

\bibitem{Podolsky}B. Podolsky and P. Schwed, Review of a Generalized
Electrodynamics. \href{http://dx.doi.org/10.1103/RevModPhys.20.40}{Rev. Mod. Phys.}
20, 40 (1948).

\bibitem{Frenkel}J. Frenkel. 4/3 problem in classical electrodynamics.
\href{http://dx.doi.org/10.1103/PhysRevE.54.5859}{Phys. Rev. E} 54,
5859 (1996).

\bibitem{Williams}E. R. Williams, J. E. Faller, and H. A. Hill. New
Experimental Test of Coulomb\textquoteright s Law: A Laboratory Upper
Limit on the Photon Rest Mass. \href{http://dx.doi.org/10.1103/PhysRevLett.26.721}{Phys. Rev. Lett.}
26, 721 (1971).\\
L. Davis, A. S. Goldhaber, and M. M. Nieto. Limit on the Photon Mass
Deduced from Pioneer-10 Observations of Jupiter\textquoteright s Magnetic
Field. \href{http://dx.doi.org/10.1103/PhysRevLett.35.1402}{Phys. Rev. Lett.}
35, 1402 (1975).

\bibitem{Luo}Jun Luo, Liang-Cheng Tu, Zhong-Kun Hu, and En-Jie Luan.
New Experimental Limit on the Photon Rest Mass with a Rotating Torsion
Balance. \href{http://dx.doi.org/10.1103/PhysRevLett.90.081801}{Phys. Rev. Lett.}
90, 081801 (2003).

\bibitem{rel.dyn.}P. A. M. Dirac, Forms of Relativistic Dynamics,
\href{http://dx.doi.org/10.1103/RevModPhys.21.392}{Rev. Mod. Phys.}
21 (1949) 392.\\
 B. L. G. Bakker, Forms of Relativistic Dynamics, Lecture Notes in
Physics 572, Springer, New York, 2001.

\bibitem{Huszar}M. Huszár, Light front quantization by Dirac's method,
J. Phys. A: \href{http://dx.doi.org/10.1088/0305-4470/9/8/026}{Math. Gen.}
9 (1976) 1359.

\bibitem{Steinhardt}P. J. Steinhardt, Problems of quantization in
the infinite momentum frame, \href{http://dx.doi.org/10.1016/0003-4916(80)90327-9}{Ann. Phys}
128 (1980) 425.

\bibitem{Galvao}C. A. P. Galvão and B. M. Pimentel. The canonical
structure of Podolsky generalized electrodynamics. \href{http://dx.doi.org/10.1139/p88-075}{Can. J. Phys.}
66 (1988) 460\textendash 466.

\bibitem{Pod1}M. C. Bertin, B. M. Pimentel, and G. E. R. Zambrano,
The canonical structure of Podolsky's generalized electrodynamics
on the null-plane, \href{http://dx.doi.org/10.1063/1.3653510}{J. Math. Phys.}
52 (2011) 102902.

\bibitem{Castellani}L. Castellani, Symmetries in constrained hamiltonian
systems, \href{http://dx.doi.org/10.1016/0003-4916(82)90031-8}{Ann. Phys.}
143, 357 (1982).

\bibitem{Weyl}H. Weyl, Geodesic Fields in the Calculus of Variation
for Multiple Integrals, \href{https://www.jstor.org/stable/1968645?seq=1\#page_scan_tab_contents}{Annals of Mathematics},
Second Series 36, 607 (1935).

\bibitem{key-2}R. R. Cuzinatto, C. A. M. De Melo, L. G. Medeiros,
and P. J. Pompeia, How Can One Probe Prodolsky Electrodynamics? \href{http://dx.doi.org/10.1142/s0217751x11053961}{Int. J. Mod. Phys. A}
26, 3641 (2011).\end{thebibliography}
\end{document}